\documentclass[12pt]{article}

\usepackage{graphicx}
\usepackage{amsmath}
\usepackage{amssymb}
\usepackage{amsthm}
\usepackage{bm}
\numberwithin{equation}{section}

\newcommand{\Tr}{{\rm Tr}}

\makeatletter

\begin{document}
\baselineskip=16pt plus 0.2pt minus 0.1pt
\renewcommand{\thefootnote}{\fnsymbol{footnote}}

\begin{titlepage}
\begin{flushright}
MISC-2011-05
\end{flushright}
\begin{center}
{\Large\bf Heterotic $E_6$ GUTs and Partition Functions}\\
\bigskip\bigskip\bigskip
{\large  
$^a$Motoharu Ito\footnote{mot@eken.phys.nagoya-u.ac.jp},
$^a$Shogo Kuwakino\footnote{skuwa@eken.phys.nagoya-u.ac.jp},
$^{ab}$Nobuhiro Maekawa\footnote{maekawa@eken.phys.nagoya-u.ac.jp},\\
$^{bcd}$Sanefumi Moriyama\footnote{moriyama@math.nagoya-u.ac.jp},
$^e$Keijiro Takahashi\footnote{takahashi-2nyh@jp.nomura.com},
$^a$Kazuaki Takei\footnote{takei@eken.phys.nagoya-u.ac.jp},\\
$^{af}$Shunsuke Teraguchi\footnote{teraguch@ifrec.osaka-u.ac.jp},
$^{ag}$Toshifumi Yamashita\footnote{yamasita@eken.phys.nagoya-u.ac.jp}
}\\
\bigskip\bigskip
{\it $^a$Department of Physics, Nagoya University,
Nagoya 464-8602, Japan}\\
{\it $^b$Kobayashi Maskawa Institute, Nagoya University,
Nagoya 464-8602, Japan}\\
{\it $^c$Graduate School of Mathematics, Nagoya University,
Nagoya 464-8602, Japan}\\
{\it $^d$Yukawa Institute for Theoretical Physics, Kyoto University,
Kyoto 606-8502, Japan}\\
{\it $^e$Department of Electrophysics, National Chiao-Tung University,
Hsinchu, Taiwan, R.O.C.}\\
{\it $^f$WPI Immunology Frontier Research Center, Osaka University,
Osaka 565-0871, Japan}\\
{\it $^g$MISC, Kyoto Sangyo University,
Kyoto 603-8555, Japan}
\end{center}
\begin{abstract}
\normalsize
The $E_6$ grand unified theory is an attractive candidate intermediate
theory between the standard model and string theory.
However, only one $E_6$ grand unified model with three generations and
at least one adjoint Higgs field has been derived from string theory
in the literature, and this model is phenomenologically
unsatisfactory.
Recently, in arXiv:1012.1690, we have constructed two new such $E_6$
grand unified models in heterotic asymmetric orbifolds.
Although our new models themselves cannot resolve the unsatisfactory
point in the previous model, our discovery raises hopes that one can
construct many other such models in this framework and find better
models.
Here, by giving partition functions explicitly, we explain the details
of our construction.
Utilizing the lattice engineering technique and the diagonal embedding
method, we can construct models systematically.
We hope that these techniques and the details of our construction will
lead to more phenomenologically desirable models.
\end{abstract}
\end{titlepage}

\setcounter{footnote}{0}

\renewcommand{\thefootnote}{\arabic{footnote}}

\section{Introduction}

The standard model is unsatisfactory from an aesthetic viewpoint.
Even though the model has very good agreement with experimental results, the
matter content appears to be too complicated to be the most fundamental
structure of the universe.
This unsatisfactory point can be stated in a more scientific way.
Despite the complicated matter content in the model, after summing over
their contributions, the quantum anomaly cancels miraculously.
This miraculous cancellation has to be explained.
Considering that the periodic table of the chemical
elements finally led to the discovery of the subatomic structure and
that the representation-theoretical diagrams of hadrons
led to the discovery of quarks, it is natural to expect a more
fundamental structure beyond the standard model.

A fundamental structure was proposed as the grand unified theory
(GUT) \cite{GUT,SUSYGUT}.
For example, if we regard the standard model gauge group
$SU(3)_C\times SU(2)_L\times U(1)_Y$ as $E_3\times U(1)$ and continue
the unification in the exceptional algebras of the $E$-series, we find
that the matter content of each generation is unified into a few
multiplets or a single multiplet in $E_4=SU(5)$ or $E_5=SO(10)$,
respectively.
Furthermore, in the $E_6$ unification \cite{E6}, the seemingly redundant 
fields in the fundamental representation ${\bf 27}$ are useful for
explaining the hierarchical structure of the quark-lepton masses and mixings 
in the standard model in a simple way \cite{Etwist,horizontal}\footnote{ 
In addition, there are no chiral exotics with respect to the standard model 
gauge group in $E_6$ models.}.
Reviewing all of these beautiful unifications, 
it seems reasonable to anticipate the emergence
of an $E$-series structure independent of the details of models.
It is also natural to require four-dimensional ${\mathcal N}=1$
supersymmetry (SUSY) to enforce the gauge-coupling unification and
adjoint Higgs fields to enable symmetry breaking in this context.

String theory is another candidate fundamental structure which also
unifies gravity.
Among the known frameworks for phenomenological studies of string 
theory \cite{heteroph,D-brane,F-GUT}, heterotic string
theory \cite{hetero} matches particularly well with the $E$-series and has a
well-defined Lagrangian description.
Therefore, aside from any further specific phenomenological requirements,
it is interesting to ask whether we can find unified models with
\begin{itemize}
\item an $E_6$ unification group,
\item Higgs fields in the adjoint representation,
\item three generations,
\item four-dimensional ${\mathcal N}=1$ SUSY,
\end{itemize}
from heterotic string theory.

In the construction of phenomenological models in heterotic string 
theory, compactifications on geometric Calabi-Yau manifolds or 
symmetric orbifolds \cite{orbifold} are usually utilized.
In symmetric orbifolds, the standard ten-dimensional heterotic string 
theory with an extra left-moving rank 16 gauge group 
$E_8\times E_8$ or $SO(32)$ is compactified on a six-dimensional 
orbifold, which is defined by a six-dimensional torus  divided by its 
rotation symmetries. 
The orbifold action $\theta$ is basically defined for the target 
space, $\theta: x\mapsto\theta x$, with a clear geometric picture.
In terms of its world-sheet theory, the action is common for the 
left-movers $X_L(z)$ and right-movers $X_R(\overline z)$ of the 
string coordinates.
In the case of heterotic string where fermions are only right-moving,
the orbifold action $\theta$ is extended to act on the left-moving
$E_8\times E_8$ or $SO(32)$ lattice
to compensate the asymmetry.
It has been found, however, that in these compactifications it is 
difficult to find unified models with the above requirements.

Considering that the orbifold action can be generalized so as to also act on 
the extra left-moving lattice, it is natural to generalize 
the orbifold action further 
so that it acts on the left-moving and right-moving lattices separately.
This type of construction is called an asymmetric orbifold \cite{asymmetric}.
Namely, in asymmetric orbifolds, the action can be defined 
independently on the left-movers and right-movers,
\begin{align}
\theta:X_L\mapsto\theta_LX_L,\quad X_R\mapsto\theta_RX_R,
\end{align}
with $\theta_L \neq \theta_R$.
In general, the starting point is not necessarily restricted to the 
$E_8 \times E_8$ or $SO(32)$ heterotic theory 
but may include heterotic theory compactified on a general Lorentzian even self-dual lattice \cite{Narain},
which combines the geometric six dimensions and the extra
left-moving 16 dimensions together from the beginning.
Since, in this paper, we describe even self-dual lattices using Lie lattices,
asymmetric orbifold actions are defined as orbifold identifications of
discrete symmetries of these Lie lattices.
Compared with symmetric orbifolds, asymmetric orbifolds offer
many more possibilities for model construction since
there are many possible even self-dual lattices and asymmetric orbifold actions,
although the consistency condition is more complicated.

A thorough study on heterotic asymmetric orbifolds in \cite{KTE6} showed 
that such construction of $E_6$ GUTs is actually possible. 
The authors of \cite{KTE6} claimed that, under the additional requirement of a hidden non-Abelian 
gauge group for SUSY breaking \cite{IzawaYanagida},
only one model with the above physical requirements exists.
Since, however, it is known that there are other possibilities for breaking the
SUSY such as that in \cite{SUSYbr}, in which our world is realized in a
metastable vacuum \cite{ISS},
the requirement of the hidden non-Abelian gauge group may be relaxed 
to construct new models. 
In addition, unfortunately there are no mechanisms to prevent the 
doublet-triplet splitting problem and the SUSY flavor/CP problem in their 
unique model in \cite{KTE6}.
Considering possible solutions to the problems utilizing additional gauge 
symmetries such as the anomalous $U(1)_A$ gauge symmetry 
\cite{anomalousU(1),anomalousU(1)ph,anomalousU(1)GUT} and $SU(2)_F$ 
family symmetry \cite{horizontal,SCPV}, it
is worth checking whether or not such additional symmetries can be realized
in these new models.

We revisited this direction in \cite{letter}, 
where we systematically translated the above four
physical requirements into a setup in string theory.
As a result, we found three models 
with one
of them having the same massless spectrum as that in \cite{KTE6} and two
of them being new.
Surprisingly, we found that one of the new models contained a further
hidden non-Abelian gauge group and was dropped from the classification
in \cite{KTE6}.
In addition to these explicitly constructed models, we stress that
the techniques used in \cite{letter} are now well established and we can construct many 
types of models at will.
Unfortunately, our new models share the same phenomenologically
unsatisfactory issues as the model in \cite{KTE6}.
Despite this, we believe that our discovery of new $E_6$ models in a
systematic way is important because it raises hopes that more
phenomenologically attractive models will be found.

In the current paper, we present details of the model construction
in \cite{letter} using partition functions in a self-contained way.
The techniques used include the lattice engineering technique \cite{LSW}, modular 
invariance with arbitrary shift actions and the diagonal embedding 
method \cite{DiagonalEmbedding} with a shift action.
As explained in \cite{letter}, various physical requirements can be 
translated into the string setup using the above techniques. 
Note that, although we have applied these techniques to $E_6$ GUT model 
construction in this study, the same techniques can be used for constructing 
models with other unified gauge symmetries such as $SO(10)$, $SU(5)$ 
and the standard model group $SU(3)_C\times SU(2)_L\times U(1)_Y$.
We shall briefly review our setup from the viewpoint of the following physical
requirements.

{\it $E_6$ unification group.}
Since we are considering the compactification of heterotic string
theory, the momentum space is quantized on a lattice.
{}From the consistency condition of the modular invariance in string
theory, the lattice is required to be \cite{Narain}
\begin{itemize}
\item even, meaning that
$({\bm e}_in^i)^2=({\bm e}_in^i)\circ({\bm e}_jn^j)\in 2{\mathbb Z}$
for $n^i\in{\mathbb Z}$, with ${\bm e}_i$ being the lattice basis and
$\circ$ being the inner product in the lattice space, and
\item self-dual, meaning that
$\{{\bm e}_in^i|n^i\in{\mathbb Z}\}
=\{\widetilde{\bm e}_im^i|m^i\in{\mathbb Z}\}$,
with $\widetilde{\bm e}_i$ being the dual lattice basis such that
${\bm e}_i\circ\widetilde{\bm e}_j=\delta_{ij}$.
\end{itemize}
According to \cite{Narain}, provided we have an extra even self-dual
(22,6)-dimensional lattice (which denotes a lattice containing a
22-dimensional left-moving lattice and a six-dimensional right-moving lattice), we
can obtain a consistent four-dimensional string theory without
considering its 10-dimensional origin.
Hence, hereafter we specify our unorbifolded theory by its
(22,6)-dimensional lattice.
Since the left-moving part of the lattice contributes to the spacetime
gauge symmetry, we have to construct an even self-dual lattice
containing $E_6$ in the left-moving part.

{\it Adjoint Higgs fields.}
In general, when heterotic string theory realizes a spacetime
gauge symmetry, the currents of the corresponding worldsheet theory
form the Kac-Moody algebra
\begin{align}
[j^a_m,j^b_n]=if^{ab}{}_cj^c_{m+n}+km\delta^{ab}\delta_{m+n,0}.
\end{align}
Here $j^a_m$ is the Kac-Moody current and $f^{ab}{}_c$ is its structure
constant.
In the above construction of even self-dual lattices, we typically find
the Kac-Moody level to be $k=1$.
It is known, however, that to obtain adjoint Higgs fields we need to
increase the Kac-Moody level to $k>1$ \cite{adjoint}.
For this purpose, we utilize the diagonal embedding method
\cite{DiagonalEmbedding}. 
Namely, we select $K$ copies of the above Kac-Moody current denoted
by $I$ with $I=1,\cdots, K$ and consider the orbifold 
action by permuting them.
Then, the remaining diagonal current after the orbifold projection 
\begin{align}
J_{\rm diag}=\sum_{I=1}^Kj_I
\end{align}
satisfies the same Kac-Moody algebra with the level $k=K$.

{\it Three generations.}
To obtain a nonvanishing generation number, we have to introduce a
shift action in addition to the permutation action to break the symmetry between
chiral and antichiral matter.
Unfortunately, it is not easy to translate the condition of three
generations into the string theory setup.
Since there is a conjecture stating that the generation number is a
multiple of the Kac-Moody level \cite{KTE6}, we choose the Kac-Moody
level to be three here.
For this purpose, our lattice has to contain a left-moving $(E_6)^3$
lattice, which does not fit the 16 extra left-moving dimensions. 
This is why we consider heterotic string theory with a Narain 
compactification \cite{Narain} instead of the standard $E_8\times E_8$ or $SO(32)$ 
heterotic string theory.

{\it ${\mathcal N}=1$ SUSY.}
To meet the requirement of ${\mathcal N}=1$ SUSY, we need a
suitable orbifold projection on the right-moving lattice.
For the right-moving $E_6$ part, a typical choice is the
${\mathbb Z}_{12}$ Coxeter element.

For the construction of the desired even self-dual lattice, 
the lattice engineering technique is useful \cite{LSW}.
This technique allows us to generate a new even self-dual lattice from
a known one.
The essence of this technique is to utilize the
fact that a lattice (say, an $A_2$ lattice) transforms oppositely under
the modular transformation compared with its complement lattice in the $E_8$
lattice (the $E_6$ lattice for the case of the above example of an $A_2$
lattice).
Using this fact, we can always replace the left-moving $A_2$ lattice
with the right-moving $[E_6]^*$ lattice and vice versa.
Here we denote the right-moving lattice with an asterisk because it
contributes to the partition function in the complex conjugate.
Using this technique, we can always obtain an even self-dual lattice containing a
left-moving $(E_6)^3$ sublattice starting from a lattice containing a
left-moving $A_2$ sublattice.
Namely, we can always replace the left-moving $A_2$ sublattice with a right-moving $[E_6]^*$ lattice
and, after subsequent decomposition into $[(A_2)^3]^*$ and further
replacements, we end up with a lattice containing $(E_6)^3$.
For example, in \cite{letter} we prepare the $E_6\times[E_6]^*$ lattice 
as a known even self-dual.
After decomposing the left-moving $E_6$ part into $(A_2)^3$, we can
replace one of the $A_2$ by $(E_6)^3$ and obtain an even self-dual
$[(A_2)^2\times(E_6)^3]\times[E_6]^*$ lattice.
Note that this technique is merely a mathematical tool for finding new even self-dual 
lattices and is unconnected with whether or not we can construct a
heterotic string theory from it.

At this stage, it may appear that the above requirements restrict possible
lattices too strongly and that there is little room to construct many 
models.
However, we can introduce further orbifold actions on the two $A_2$
lattices of $[(A_2)^2\times(E_6)^3]\times[E_6]^*$, 
which add variety to models
without changing the above properties including the modular invariance.
Therefore, we also classify all the possible shift and rotation
actions on the two $A_2$ lattices in this paper.

The outline of this paper is as follows. In the next section, we
define the lattice partition functions with a general shift action, which
are needed for the asymmetric orbifold construction of our $E_6$ models.
We also explain the lattice engineering technique and the
diagonal embedding method used to obtain adjoint Higgs fields. In section
3, we present the setup of our model construction explicitly and
classify all the possible models in this framework. In section 4, we
analyze the massless spectra of our new models with three generations
in detail. Section 5 is devoted to a summary and discussion. Other
technical details are left to the appendices, where we summarize the
partition functions of bosonic and fermionic oscillators and give a
short review on some useful decompositions of Lie lattices.

\section{Lattice partition functions}\label{LatandPF}
The one-loop partition function of closed string theory is defined by
\begin{align}
Z(\tau)=
\Tr_{\mathcal H}q^{L_0-a}{\overline q}^{\overline L_0-\overline a},
\label{partition_def}
\end{align}
for a modular parameter $\tau$ and $q=e^{2\pi i\tau}$.
Here the trace is taken over the closed string Hilbert space ${\mathcal H}$,
while $L_0$($\overline L_0$) and $a$($\overline a$) are the Virasoro zero mode
and zero-point energy of the left(right)-moving modes, respectively.
The modular transformations that identify the different moduli $\tau$
form a discrete group $PSL(2,{\mathbb Z})$ and are generated by
\begin{align}
{\mathcal T}:\tau\mapsto\tau+1,\quad{\mathcal S}:\tau\mapsto-1/\tau.
\label{modular}
\end{align}
The partition function (\ref{partition_def}) should be invariant 
under the transformations
\begin{align}
Z(\tau+1)=Z(\tau),\quad Z(-1/\tau)=Z(\tau).
\end{align}

In ${\mathbb Z}_N$ orbifold theory, the partition function is
divided into various sectors labeled by $(\alpha,\beta)$,
\begin{align}
Z(\tau)=
\frac{1}{N}\sum_{\alpha,\beta=0}^{N-1}Z\Bigl[
\begin{matrix}\alpha\\\beta\end{matrix}\Bigr](\tau),\quad
Z\Bigl[\begin{matrix}\alpha\\\beta\end{matrix}\Bigr](\tau)
=
\Tr_{{\mathcal H}_\alpha}
q^{L_0-a}{\overline q}^{\overline L_0-\overline a}\theta^\beta,
\label{defPF}
\end{align}
with $\theta$ being the orbifold action.
Here, ${\mathcal H}_\alpha$ is the Hilbert space of the $\alpha$ twisted sector.
These sectors should transform covariantly under the
modular transformation
\begin{align}
Z\Bigl[\begin{matrix}\alpha\\\beta\end{matrix}\Bigr](\tau+1)
=Z\Bigl[\begin{matrix}\alpha\\\beta+\alpha\end{matrix}\Bigr](\tau),\quad
Z\Bigl[\begin{matrix}\alpha\\\beta\end{matrix}\Bigr](-1/\tau)
=Z\Bigl[\begin{matrix}\beta\\-\alpha\end{matrix}\Bigr](\tau),
\label{orbtransf}
\end{align}
supplemented by the orbifold periodic consistency condition
\begin{align}
Z\Bigl[\begin{matrix}\alpha\\\beta\end{matrix}\Bigr](\tau)
=Z\Bigl[\begin{matrix}\alpha\\\beta+N\end{matrix}\Bigr](\tau).
\label{orbperiod}
\end{align}
In the following, we divide the partition function into several
parts originating from the fermions, the bosonic oscillators and the
zero-mode momentum states on a lattice.
In studying each contribution, we typically define a partition
function for arbitrary $(\alpha,\beta)$ so that it satisfies the
modular transformation \eqref{orbtransf}.
We only require the orbifold periodic condition \eqref{orbperiod} of
each part to hold up to a phase, since various phases may cancel
each other after summing over all the contributions.

In the asymmetric orbifold construction, the main complication arises 
from the lattice part
\begin{align}
\Theta^{\rm L}(\tau) =  \sum_{(p_L,p_R)\in\Gamma}q^{p_L^2/2}\overline q^{p_R^2/2},
\label{latPF}
\end{align}
where $\Gamma$ denotes the lattice.
Therefore, in this paper, we focus on the lattice partition function and give
the definition of the fermion and bosonic oscillator partition functions 
in appendix \ref{osc_pf}.

\subsection{Lattices}\label{latticesubsec}
To obtain the four-dimensional spacetime in heterotic string theory 
we have to compactify the $(22,6)$-dimensional spacetime.
After compactification, the momenta are quantized and reside on
a $(22,6)$-dimensional Lorentzian lattice.
A lattice is a set of points that are generated by a set of basis
vectors ${\bm e}_i$ with integral coefficients:
$\{{\bm e}_in^i|n^i\in{\mathbb Z}\}$.
A lattice is even when it satisfies
$({\bm e}_in^i)^2\in 2{\mathbb Z}$.
The dual lattice is a lattice
$\{\widetilde{\bm e}_im^i|m^i\in{\mathbb Z}\}$ generated by
the dual basis $\widetilde{\bm e}_i$ of the original lattice, which
satisfies ${\bm e}_i\circ\widetilde{\bm e}_j=\delta_{ij}$.
A lattice is self-dual when the dual lattice is exactly the same as
the original one.
For the modular invariance of string theory, we require the
lattice to be even and self-dual. 
In heterotic string theory, the left-moving part of the Lie
lattice is responsible for the spacetime Lie-algebraic gauge symmetry.
In a Lie algebra, the root lattice is generated by the simple roots 
${\bm\alpha}_i$, and it is known to be an even lattice for the case of
a simply laced Lie algebra.
Therefore, we are especially interested in simply laced Lie
algebras.
A weight lattice is generated by the fundamental weights
${\bm\omega}_i$ satisfying
${\bm\alpha}_i\circ{\bm\omega}_j=\delta_{ij}$.
In other words, a weight lattice is the dual lattice of a root
lattice and, in fact, a root lattice is a sublattice of its weight lattice.

Since we have already chosen even lattices, it is desirable to know how close they are to being
self-dual lattices and how we can generate even self-dual lattices from this knowledge.
An efficient way to study the above questions is 
to use conjugacy classes.
Conjugacy classes can be defined by
identifying points of the weight lattice, whose difference resides in the root
lattice.
For $E_6$ and $A_2$, which are our main concern in this paper, 
the conjugacy classes are isomorphic to ${\mathbb Z}_3$, with
the generator being the weight vector of the fundamental
representation with the minimal dimension.
This means that the conjugacy classes of $E_6$ (or $A_2$, respectively)
have three elements, namely, the root lattice, that
shifted by the weight of the fundamental representation
${\bf 27}$ (or ${\bf 3}$) and that shifted by the weight of
the antifundamental representation $\overline{\bf 27}$ (or
$\overline{\bf 3}$).
These elements have the same additive
structure as the additive group $\{0,1,2\}$ mod 3.

In the following, we explain
various techniques used in constructing our $E_6$ unified models
\cite{letter},
including the lattice engineering technique, orbifolds with general
shift actions and permutation with a specific shift.
In appendix \ref{lattice_decomp}, we summarize some useful 
decompositions of Lie lattices used in our analysis
in terms of their conjugacy classes.

\subsection{Lattice engineering technique}
As mentioned in the previous subsection, modular
invariance requires the momentum lattice to be even and self-dual.
Therefore, our starting point in studying heterotic string theory
is to search for an even self-dual lattice with the desired properties.
In this subsection, we explain the lattice engineering
technique \cite{LSW}, using which we can construct a new even
self-dual lattice out of a given one with different dimensionality.
In fact, it turns out that this technique is particularly useful for constructing models with $E_6$ gauge symmetry
 with Kac-Moody level 3,
where we need an even
self-dual lattice containing $(E_6)^3$.

Here we study the $E_6$ lattice and $A_2$ lattice.
The partition functions of the $A_2$ lattice and the lattice shifted 
by the fundamental weight (denoted as $A(\tau)$ and $a(\tau)$, respectively)
can be expressed in terms of the standard theta function:
\begin{align}
&A(\tau)=\vartheta\Bigl[\begin{matrix}0\\0\end{matrix}\Bigr](2\tau)
\vartheta\Bigl[\begin{matrix}0\\0\end{matrix}\Bigr](6\tau)
+\vartheta\Bigl[\begin{matrix}1/2\\0\end{matrix}\Bigr](2\tau)
\vartheta\Bigl[\begin{matrix}1/2\\0\end{matrix}\Bigr](6\tau),
\nonumber\\
&a(\tau)=\vartheta\Bigl[\begin{matrix}0\\0\end{matrix}\Bigr](2\tau)
\vartheta\Bigl[\begin{matrix}1/3\\0\end{matrix}\Bigr](6\tau)
+\vartheta\Bigl[\begin{matrix}1/2\\0\end{matrix}\Bigr](2\tau)
\vartheta\Bigl[\begin{matrix}5/6\\0\end{matrix}\Bigr](6\tau).
\end{align}
Note that the root lattice shifted by the antifundamental weight is 
actually the same as that shifted by 
twice the fundamental weight and takes the same partition function, $a(\tau)$. 
From the decompositions $E_8\to E_6\times A_2$ and
$E_6\to(A_2)^3$ (reviewed in \eqref{E8E6A2} and \eqref{E6A2^3}), 
it is not difficult to obtain the relations
\begin{align}
\Theta^{\rm L}_{E_8}(\tau)=E(\tau)A(\tau)+2e(\tau)a(\tau)
\label{EA2ea}
\end{align}
and
\begin{align}
E(\tau)=A(\tau)^3+2a(\tau)^3,\quad
e(\tau)=3A(\tau)a(\tau)^2.
\end{align}
Here, $\Theta^{\rm L}_{E_8}(\tau)$ is the partition function of
the $E_8$ root lattice, and $E(\tau)$ and $e(\tau)$ are
the partition function of the $E_6$ root lattice and that of the root
lattice shifted by its fundamental (or antifundamental) weight, respectively.
Under the modular transformations, these partition functions transform as
\begin{align}
A(\tau+1)&=A(\tau),&
A(-1/\tau)&=\sqrt{-i\tau}^2\frac{1}{\sqrt{3}}(A+2a)(\tau),\nonumber\\
a(\tau+1)&=\omega a(\tau),&
a(-1/\tau)&=\sqrt{-i\tau}^2\frac{1}{\sqrt{3}}(A-a)(\tau),\nonumber\\
E(\tau+1)&=E(\tau),&
E(-1/\tau)&=\sqrt{-i\tau}^6\frac{1}{\sqrt{3}}(E+2e)(\tau),\nonumber\\
e(\tau+1)&=\omega^{-1}e(\tau),&
e(-1/\tau)&=\sqrt{-i\tau}^6\frac{1}{\sqrt{3}}(E-e)(\tau),
\label{modularEeAa}
\end{align}
with $\omega=\exp(2\pi i/3)$.
If we respectively define $A_k(\tau)$ and $E_k(\tau)$ to be the $A_2$ and $E_6$
partition functions shifted by $k$-multiples of their fundamental
weights,
\begin{align}
A_k(\tau)=\bigl(A(\tau),a(\tau),a(\tau)\bigr),\quad
E_k(\tau)=\bigl(E(\tau),e(\tau),e(\tau)\bigr),
\end{align}
the above modular transformation can be expressed as
\begin{align}
A_k(\tau+1)&=\omega^{k^2}A_k(\tau),&
A_k(-1/\tau)&=\sqrt{-i\tau}^2
\frac{1}{\sqrt{3}}\sum_{l=0}^2\omega^{-kl}A_l(\tau),\nonumber\\
E_k(\tau+1)&=\omega^{-k^2}E_k(\tau),&
E_k(-1/\tau)&=\sqrt{-i\tau}^6
\frac{1}{\sqrt{3}}\sum_{l=0}^2\omega^{kl}E_l(\tau).
\label{AEmodular}
\end{align}
One might notice that $A_k$ and $E_k$ transform
oppositely under the modular transformation.
This property becomes even more manifest if we rewrite the 
decomposition \eqref{EA2ea} of the $E_8$  partition function, 
which is invariant under modular transformations, 
in terms of $A_k$ and $E_k$ as
\begin{align}
\Theta^{\rm L}_{E_8}(\tau)=\sum_{k=0}^2A_k(\tau)E_k(\tau).
\label{E8toE6A2}
\end{align}
Thus,
the partition function of a lattice (say, an $A_2$ lattice) transforms
oppositely compared with the partition function of its complement lattice
in the $E_8$ lattice ($E_6$ lattice for the case of the above example of an
$A_2$ lattice).
Since the left-moving and right-moving lattice partition functions
also transform oppositely, we can construct
a new lattice without changing its modular transformation property by
replacing the left-moving $A_2$ lattice with the right-moving
$[E_6]^*$ lattice and vice versa.
Thus, one can always generate new even self-dual lattices from known ones.
This is the lattice engineering technique \cite{LSW}.

We show some examples of the lattice engineering technique.
For this purpose, we rewrite \eqref{E8toE6A2} as 
\begin{align}
\Theta^{\rm L}_{E_8}(\tau)=\sum_{(k_A,k_E)\in\Pi_{E_8}}A_{k_A}(\tau)E_{k_E}(\tau),
\qquad
\Pi_{E_8}=\{(0,0),(1,1),(2,2)\}
\label{DefPi}
\end{align}
and use the set of conjugacy classes $\Pi_{E_8}$ as the definition of the lattice.
In the current case, the conjugacy classes do not change upon the replacement of 
the lattices in the lattice engineering technique.
For example, if we replace the left-moving $E_6$ lattice with the right-moving $[A_2]^*$ lattice,
the resulting lattice is the $(2,2)$-dimensional even self-dual lattice $A_2 \times [A_2]^*$.
Here, the corresponding partition function and conjugacy classes of the 
$A_2 \times [A_2]^*$ lattice are respectively given as
\begin{align}
\Theta^{\rm L}_{A_2 \times [A_2]^*}(\tau)
=\sum_{(k_A,k_{A^*})\in\Pi_{A_2 \times [A_2]^*}}
A_{k_A}(\tau) \left[ A_{k_{A^*}}(\tau) \right]^*,
\ \ \ 
\Pi_{A_2 \times [A_2]^*}=\{(0,0),(1,1),(2,2)\}.
\label{DefPiA2A2bar}
\end{align}
On the other hand, if we consider the replacement of the left-moving $A_2$ lattice 
with the right-moving $[E_6]^*$ lattice, we obtain 
the $(6,6)$-dimensional even self-dual lattice $E_6 \times [E_6]^*$ 
with the partition function and conjugacy classes 
\begin{align}
\Theta^{\rm L}_{E_6 \times [E_6]^*}(\tau)=
\sum_{(k_E,k_{E^*})\in\Pi_{E_6 \times [E_6]^*}}
E_{k_E}(\tau) \left[ E_{k_{E^*}}(\tau) \right]^*,
\ \ \
\Pi_{E_6 \times [E_6]^*}=\{(0,0),(1,1),(2,2)\}.
\label{DefPiE6E6bar}
\end{align}
Furthermore, we can use this technique iteratively by considering
the decomposition of an even self-dual lattice into various sublattices and further replacements.
Note that after the decomposition,
conjugacy classes are expressed in terms of the corresponding subalgebras.

For the construction of our model, we employ the following subsequent decompositions and 
replacements:
\begin{align}
&E_8\to E_6\times A_2\to E_6\times[E_6]^*
\to(A_2)^3\times[E_6]^*
\to(A_2)^2\times[E_6\times E_6]^*\nonumber\\
&\qquad\to(A_2)^2\times[(A_2)^3\times E_6]^*
\to[(A_2)^2\times(E_6)^3]\times[E_6]^*.
\label{lattice}
\end{align}
The corresponding conjugacy classes for these processes will be given later 
in subsection \ref{sec:Unorbifolded theory}.
Similarly, provided we have an even self-dual lattice containing
$A_2$, we can always construct another even self-dual lattice
containing $(E_6)^3$.
For example, we can construct the $(18,2)$-dimensional even self-dual
lattice $(E_6)^3\times[A_2]^*$ out of $A_2\times[A_2]^*$ and the
$(20,4)$-dimensional even self-dual lattice
$[\widetilde A_2\times(E_6)^3]\times[D_4]^*$ out of
$D_4\times[D_4]^*$ using the decomposition \eqref{D4A2A2}.

Note that although the dimensionality of even self-dual lattices varies in the
lattice engineering technique,
this is unrelated to the dimensionality of the string theory.
We are simply employing the resultant $(22,6)$-dimensional even self-dual lattice with
the desired properties after performing lattice engineering
for our $(26,10)$-dimensional heterotic string theory.

\subsection{General shift actions}\label{shiftaction}

In this subsection, we introduce the general shift actions for
the two $A_2$ lattices in \eqref{lattice}.
In fact, it will turn out that the modular transformation property does not change
even if we introduce shift actions.
Therefore, we can treat models with and without shifts equally.

In the following, we begin with the case without shift actions, 
and then we generalize to the case with shift actions with the help of
the generalized theta functions that we introduce subsequently.
For the purpose of explanation, we consider the
$A_2\times[A_2]^*$ lattice with the conjugacy classes
$\oplus_{k=0}^2k(1,1)$ and perform an orbifold projection on the
right-moving part $[A_2]^*$ by the ${\mathbb Z}_3$ twist with the
rotation angle $2\pi\times 1/3$.

\subsubsection{Partition functions without shift actions}
Here we study the partition function without shifts.
Since only the origin is invariant under a nontrivial twist, among
the conjugacy classes, only $(0,0)$, which contains the origin of $[A_2]^*$,
remains.
Thus, our lattice partition functions in the untwisted sectors are simply
$A(\tau)$ (up to a phase ambiguity). 
Therefore, we shall define the $A_2$ lattice partition function 
for each sector of the orbifold theory, 
$A\Bigl[\begin{matrix}\alpha\\\beta\end{matrix}\Bigr](\tau)$, as
\begin{align}
\left[\begin{matrix}
&A\Bigl[\begin{matrix}1\\0\end{matrix}\Bigr](\tau)
&A\Bigl[\begin{matrix}2\\0\end{matrix}\Bigr](\tau)\\
A\Bigl[\begin{matrix}0\\1\end{matrix}\Bigr](\tau)
&A\Bigl[\begin{matrix}1\\1\end{matrix}\Bigr](\tau)
&A\Bigl[\begin{matrix}2\\1\end{matrix}\Bigr](\tau)\\
A\Bigl[\begin{matrix}0\\2\end{matrix}\Bigr](\tau)
&A\Bigl[\begin{matrix}1\\2\end{matrix}\Bigr](\tau)
&A\Bigl[\begin{matrix}2\\2\end{matrix}\Bigr](\tau)
\end{matrix}\right]
=\left[\begin{matrix}
&\displaystyle\frac{-i}{\sqrt{3}}(A+2a)(\tau)
&\displaystyle\frac{i}{\sqrt{3}}(A+2a)(\tau)\\[10pt]
A(\tau)
&\displaystyle\frac{-i}{\sqrt{3}}(A+2\omega a)(\tau)
&\displaystyle\frac{i}{\sqrt{3}}(A+2\omega^2a)(\tau)\\[10pt]
-A(\tau)
&\displaystyle\frac{-i}{\sqrt{3}}(A+2\omega^2a)(\tau)
&\displaystyle\frac{i}{\sqrt{3}}(A+2\omega a)(\tau)
\end{matrix}\right]
\label{A2}
\end{align}
(with periodic conditions in $\alpha$ and $\beta$),
shown compactly in a matrix form, where the first entry corresponding
to the original unorbifolded theory is omitted because 
we do not need it in our later application. 
Using \eqref{modularEeAa}, one can verify that this partition function has 
the desirable modular transformation property
\begin{align}
A\Bigl[\begin{matrix}\alpha\\\beta\end{matrix}\Bigr](\tau+1)
=A\Bigl[\begin{matrix}\alpha\\\beta+\alpha\end{matrix}\Bigr](\tau),
\quad
A\Bigl[\begin{matrix}\alpha\\\beta\end{matrix}\Bigr](-1/\tau)
=\sqrt{-i\tau}^2i
A\Bigl[\begin{matrix}\beta\\-\alpha\end{matrix}\Bigr](\tau).
\label{noshift}
\end{align}
Note that, in our convention, the ${\mathcal S}$-transformation of the partition
function of the twisted boson contains an extra phase $i$ for every
complex dimension as in \eqref{i}.
Therefore, we have defined the lattice partition
function so that its ${\mathcal S}$-transformation also acquires the
same phase $i$.

\subsubsection{Generalized theta function}
Before considering partition functions of the $A_2$ lattice with general 
shift actions, let us define the generalized theta function for a lattice 
whose metric matrix is given by $M_{ij}={\bm e}_i\circ{\bm e}_j$ as
\begin{align}
\vartheta_M
\Bigl[\begin{matrix}\vec\alpha\\\vec\beta\end{matrix}\Bigr]
=\sum_{\vec n}
\exp\bigl(-\pi(\vec n+\vec\alpha)\cdot(-i\tau M)(\vec n+\vec\alpha)
+2\pi i(\vec n+\vec\alpha)\cdot M\vec\beta\bigr).
\label{generalized}
\end{align}
The first term in the exponent is the
square of the length of the lattice state ${\bm e}_i(n+\alpha)^i$ 
shifted by ${\bm e}_i\alpha^i$ from the original lattice,
{\it i.e.}, $(\vec n+\vec\alpha)\cdot M(\vec n+\vec\alpha)
=({\bm e}_i(n+\alpha)^i)^2$,
while the second term is the phase of the state obtained by the inner
product with ${\bm e}_j\beta^j$,
$(\vec n+\vec\alpha)\cdot M\vec\beta
=\bm e_i(n+\alpha)^i\circ\bm e_j\beta^j$.
If the matrix $M$ is the Cartan matrix $C$ of a
simply laced Lie algebra and $\vec\alpha=\vec\beta=\vec 0$, 
the lattice becomes the root lattice of the
corresponding simply laced Lie algebra, and the even condition
$\vec n\cdot C\vec n\in 2{\mathbb Z}$ is automatically satisfied because of the
property of the Cartan matrix,
$C_{ii}\in 2{\mathbb Z}$ and $C_{ij}=C_{ji}\in{\mathbb Z}$ $(i\ne j)$. 
For example, if we choose the Cartan matrix of $A_2$ and $E_6$, the
generalized partition functions respectively reduce to $A(\tau)$ and
$E(\tau)$ defined in the previous subsection (or $a(\tau)$ and
$e(\tau)$ in the case that a shift by the fundamental weight is introduced).
We can prove the following modular transformation rule by using the
Poisson resummation formula:
\begin{align}
\vartheta_C
\Bigl[\begin{matrix}\vec\alpha\\\vec\beta\end{matrix}\Bigr](\tau+1)
&=e^{-\pi i\vec\alpha\cdot C\vec\alpha}
\vartheta_C
\Bigl[\begin{matrix}\vec\alpha\\\vec\beta+\vec\alpha\end{matrix}\Bigr]
(\tau),
\label{Ttransf}\\
\vartheta_C
\Bigl[\begin{matrix}\vec\alpha\\\vec\beta\end{matrix}\Bigr](-1/\tau)
&=\frac{\sqrt{-i\tau}^{\dim C}}{\sqrt{\det C}}
e^{2\pi i\vec\alpha\cdot C\vec\beta}
\vartheta_{C^{-1}}
\Bigl[\begin{matrix}C\vec\beta\\-C\vec\alpha\end{matrix}\Bigr](\tau).
\label{Stransf}
\end{align}
Here, in $\vartheta_{C^{-1}}$, the sum is taken over the weight lattice. 
Note that, after the
${\mathcal S}$-transformation, the phase assignment $\vec\beta$ is
mapped into a shift of the lattice.
In particular, if we set $\vec\alpha=\vec 0$, the momentum
lattice after the transformation is given by the weight lattice shifted
by $\vec\beta$.

Since the root lattice is a sublattice of the weight lattice, the
weight lattice can be decomposed into conjugacy classes.
The conjugacy classes of our simply laced Lie lattices are given in
Table \ref{conjugacy}\footnote{
We can generalize the following formulas in the text to the case of $D_r$
with even $r$, where we need two generators for
the conjugacy classes ${\mathbb Z}_2\times{\mathbb Z}_2$. 
For simplicity, however, we focus on the case 
that the conjugacy classes are generated by a single generator.}.
\begin{table}
\begin{center}
\begin{tabular}{c||c|c}
&{\rm conjugacy classes}&{\rm order} $(\det C)$\\\hline\hline
$A_r$&${\mathbb Z}_{r+1}$&$r+1$\\\hline
$D_r$&${\mathbb Z}_2\times{\mathbb Z}_2$ for $r\in 2{\mathbb Z}$,
${\mathbb Z}_4$ for $r\in 2{\mathbb Z}+1$&$4$\\\hline
$E_r$&${\mathbb Z}_{9-r}$&$9-r$
\end{tabular}
\end{center}
\caption{Conjugacy classes of various simply laced Lie lattices.}
\label{conjugacy}
\end{table}
Hereafter, we denote the generator of the conjugacy classes simply by
$\bm\omega$ without the index of the fundamental weight
$\bm\omega_i$.
We also denote the corresponding column of $\bm\omega$ in a
quadratic-form matrix (or the inverse of the Cartan matrix) as
$\vec F$ with its components on the diagonal line $f$, while the number
of conjugacy classes is given by $\det C$ of the Cartan
matrix $C$.
As the weight lattice is decomposed into several conjugacy classes of
the root lattice
\begin{align}
\{\vec m\cdot\vec{\bm\omega}|\vec m\in{\mathbb Z}^{\dim C}\}
=\oplus_{k=0}^{\det C-1}
\{\vec n\cdot\vec{\bm\alpha}+k\bm{\omega}
|\vec n\in{\mathbb Z}^{\dim C}\},
\end{align}
so is the shifted weight lattice
\begin{align}
\{(\vec m+\vec\alpha)\cdot\vec{\bm\omega}
|\vec m\in{\mathbb Z}^{\dim C}\}
=\oplus_{k=0}^{\det C-1}
\{(\vec n+C^{-1}\vec\alpha)\cdot\vec{\bm\alpha}
+k\bm{\omega}|\vec n\in{\mathbb Z}^{\dim C}\}.
\end{align}
Using this decomposition, we obtain the corresponding decomposition formula
of the generalized theta function as
\begin{align}
\vartheta_{C^{-1}}
\Bigl[\begin{matrix}\vec\alpha\\\vec\beta\end{matrix}\Bigr]
=\sum_{k=0}^{\det C-1}\vartheta_C
\Bigl[\begin{matrix}C^{-1}\vec\alpha+k\vec F\\
C^{-1}\vec\beta\end{matrix}\Bigr].
\label{decomposition}
\end{align}
Then, the ${\mathcal S}$-transformation \eqref{Stransf} can be
rewritten as
\begin{align}
\vartheta_C
\Bigl[\begin{matrix}\vec\alpha\\\vec\beta\end{matrix}\Bigr](-1/\tau)
&=\frac{\sqrt{-i\tau}^{\dim C}}{\sqrt{\det C}}
e^{2\pi i\vec\alpha\cdot C\vec\beta}
\sum_{k=0}^{\det C-1}\vartheta_C
\Bigl[\begin{matrix}\vec\beta+k\vec F\\
-\vec\alpha\end{matrix}\Bigr](\tau).
\label{Stransfdecomp}
\end{align}
Note also that it satisfies the integer shift formula $(\vec n\in{\mathbb Z}^{\dim C})$
\begin{align}
\vartheta_C\Bigl[\begin{matrix}\vec\alpha\\
\vec\beta+C^{-1}\vec n\end{matrix}\Bigr]
=e^{2\pi i\vec n\cdot\vec\alpha}
\vartheta_C\Bigl[\begin{matrix}\vec\alpha\\
\vec\beta\end{matrix}\Bigr].
\label{integershift}
\end{align}

\subsubsection{Partition functions with general shift actions}\label{Sec:w/GeneralShift}
We now turn to partition functions with shift actions.
As a generalization of \eqref{A2}, 
a partition function with general shift actions $\vec l$
can be defined as
\begin{align}
A^{\vec l}\Bigl[\begin{matrix}\alpha\\\beta\end{matrix}\Bigr]
=\varphi_A
\Bigl[\begin{matrix}(\alpha)_{\det C}\\(\beta)_{\det C}\end{matrix}\Bigr]
\widehat A^{\vec l}\Bigl[\begin{matrix}\alpha\\\beta\end{matrix}\Bigr]
\label{shift}
\end{align}
for various
$(\alpha,\beta)$ sectors with $\mbox{GCD}(\alpha,\beta)\ne 0$ mod
$\det C$.
Here we have divided the modular covariant partition functions $A^{\vec l}$
into a product of the numerical factors $\varphi_A$ and the physical
partition functions $\widehat A^{\vec l}$.
Each numerical factor $\varphi_A$ is given by
\begin{align}
\varphi_A
\Bigl[\begin{matrix}(\alpha)_{\det C}\\(\beta)_{\det C}\end{matrix}\Bigr]
=\begin{cases}
\varphi\bigl((\beta)_{\det C}\bigr)
&\mbox{  for  }(\alpha)_{\det C}=0,(\beta)_{\det C}\ne 0,\\[6pt]
\displaystyle
\frac{(-i)^{r/2}\varphi\bigl((\alpha)_{\det C}\bigr)}{\sqrt{\det C}}
&\mbox{  for  }(\alpha)_{\det C}\ne 0,
\end{cases}
\end{align}
where
the symbol $(n)_m$ is defined as $n$ mod $m$.
The function $\varphi$, which takes the values $\pm 1$, is introduced so
that, as discussed later, an elegant modular transformation property \eqref{A2shift} is
obtained.
For the case of the $A_2$ lattice, the explicit form of $\varphi$ will
be given later in \eqref{phi}.
On the other hand, the physical part $\widehat A^{\vec l}$ is given by
\begin{align}
\widehat A^{\vec l}
\Bigl[\begin{matrix}\alpha\\\beta\end{matrix}\Bigr]
=\begin{cases}A^{\vec l}_0
\Bigl[\begin{matrix}\alpha\\\beta\end{matrix}\Bigr]
&\mbox{  for  }(\alpha)_{\det C}=0,(\beta)_{\det C}\ne 0,\\
\displaystyle\sum_{k=0}^{\det C-1}e^{\pi i\alpha^{-1}\beta fk^2}A^{\vec l}_k
\Bigl[\begin{matrix}\alpha\\\beta\end{matrix}\Bigr]
&\mbox{  for  }(\alpha)_{\det C}\ne 0,
\end{cases}
\label{PhysA}
\end{align}
with $A^{\vec l}_k$ further defined using the generalized theta function as
\begin{align}
A^{\vec l}_k\Bigl[\begin{matrix}\alpha\\\beta\end{matrix}\Bigr]
=e^{-\pi i\alpha\beta\vec l\cdot C\vec l}
\vartheta_C\Bigl[\begin{matrix}\alpha\vec l+k\vec F\\
\beta\vec l\end{matrix}\Bigr],
\label{PhysAdef}
\end{align}
where $\alpha^{-1}$ denotes an integer that satisfies 
$\alpha^{-1}\alpha=1$ mod $\det C$.
For our application to the $A_2$ lattice, various Lie algebraic
quantities are given by
\begin{align}
r=2,\quad
C=\begin{pmatrix}2&-1\\-1&2\end{pmatrix},\quad
\vec F=\frac{1}{3}\begin{pmatrix}2\\1\end{pmatrix},\quad
f=\frac{2}{3},\quad
\varphi(1)=1,\quad\varphi(2)=-1,
\label{phi}
\end{align}
while $\alpha^{-1}$ means $1^{-1}=1$ and $2^{-1}=2$ mod 3.
Then, the numerical factors $\varphi_A$ and physical partition
functions $\widehat A^{\vec l}$ are given explicitly by
\begin{align}
\varphi_A
\Bigl[\begin{matrix}(\alpha)_3\\(\beta)_3\end{matrix}\Bigr]
&=\left[\begin{matrix}
&\displaystyle\frac{-i}{\sqrt{3}}
&\displaystyle\frac{i}{\sqrt{3}}\\[10pt]
1
&\displaystyle\frac{-i}{\sqrt{3}}
&\displaystyle\frac{i}{\sqrt{3}}\\[10pt]
-1
&\displaystyle\frac{-i}{\sqrt{3}}
&\displaystyle\frac{i}{\sqrt{3}}
\end{matrix}\right]
\end{align}
and
\begin{align}
\widehat A^{\vec l}
\Bigl[\begin{matrix}\alpha\\\beta\end{matrix}\Bigr]
&=\left[\begin{matrix}
&\displaystyle\bigl(A^{\vec l}_0+A^{\vec l}_1+A^{\vec l}_2\bigr)
\Bigl[\begin{matrix}\alpha\\\beta\end{matrix}\Bigr]
&\displaystyle\bigl(A^{\vec l}_0+A^{\vec l}_1+A^{\vec l}_2\bigr)
\Bigl[\begin{matrix}\alpha\\\beta\end{matrix}\Bigr]
\\[10pt]
A^{\vec l}_0\Bigl[\begin{matrix}\alpha\\\beta\end{matrix}\Bigr]
&\displaystyle\bigl(A^{\vec l}_0+\omega(A^{\vec l}_1+A^{\vec l}_2)\bigr)
\Bigl[\begin{matrix}\alpha\\\beta\end{matrix}\Bigr]
&\displaystyle\bigl(A^{\vec l}_0+\omega^2(A^{\vec l}_1+A^{\vec l}_2)\bigr)
\Bigl[\begin{matrix}\alpha\\\beta\end{matrix}\Bigr]
\\[10pt]
A^{\vec l}_0\Bigl[\begin{matrix}\alpha\\\beta\end{matrix}\Bigr]
&\displaystyle\bigl(A^{\vec l}_0+\omega^2(A^{\vec l}_1+A^{\vec l}_2)\bigr)
\Bigl[\begin{matrix}\alpha\\\beta\end{matrix}\Bigr]
&\displaystyle\bigl(A^{\vec l}_0+\omega(A^{\vec l}_1+A^{\vec l}_2)\bigr)
\Bigl[\begin{matrix}\alpha\\\beta\end{matrix}\Bigr]
\end{matrix}\right],
\end{align}
where the components in the matrix on the right-hand side should be
chosen as $(\alpha)_3$ and $(\beta)_3$. 
These partition functions generalize \eqref{A2}.

Again, it is not difficult to verify that this partition function
satisfies the same modular transformation property
\begin{align}
A^{\vec l}
\Bigl[\begin{matrix}\alpha\\\beta\end{matrix}\Bigr](\tau+1)
=A^{\vec l}
\Bigl[\begin{matrix}\alpha\\\beta+\alpha\end{matrix}\Bigr](\tau),
\quad
A^{\vec l}
\Bigl[\begin{matrix}\alpha\\\beta\end{matrix}\Bigr](-1/\tau)
=\sqrt{-i\tau}^2iA^{\vec l}
\Bigl[\begin{matrix}\beta\\-\alpha\end{matrix}\Bigr](\tau),
\label{A2shift}
\end{align}
using the above formulas for the generalized theta function.
Thus, the modular transformation formula \eqref{A2shift} is not
changed from \eqref{noshift} even after introducing shift actions.
This indicates that we can enhance the variety of models
by introducing shifts
which only affect the periodic relation, changing it to
\begin{align}
A^{\vec l}_k\Bigl[\begin{matrix}\alpha\\\beta+N\end{matrix}\Bigr]
=e^{\pi iN\alpha\vec l\cdot C\vec l}
A^{\vec l}_k\Bigl[\begin{matrix}\alpha\\\beta\end{matrix}\Bigr].
\label{periodA2}
\end{align}

Although for our application, we only need the $A_2$ case,
our framework here is suitable for a
general simply laced Lie lattice with $\det C$ being an odd prime
integer if we suitably define various Lie algebraic quantities.

\subsection{Permutation with a specific shift action}
\label{permutation}
In this subsection we study the partition function of
the $(E_6)^3$ sublattice with the ${\mathbb Z}_3$ orbifold action
permuting the three $E_6$ factors.
For concreteness, we consider the even self-dual lattice
$(E_6)^3\times[A_2]^*$ defined with the set of conjugacy classes
\begin{align}
&\{(0,0,0,0),(1,1,1,0),(2,2,2,0),\nonumber\\
&\;\;(0,1,2,1),(2,0,1,1),(1,2,0,1),\nonumber\\
&\;\;(0,2,1,2),(1,0,2,2),(2,1,0,2)\}
\end{align}
constructed from the lattice $A_2\times[A_2]^*$ with $\oplus_{k=0}^2k(1,1)$ 
as in \eqref{DefPiA2A2bar}. 
The lattice partition function of this even self-dual lattice is given by
\begin{align}
(E^3+2e^3)(\tau)[A(\tau)]^*+6Ee^2(\tau)[a(\tau)]^*.
\end{align}
Then, we impose the ${\mathbb Z}_3$ orbifold action by permuting the
three $E_6$ factors and simultaneously twisting the right-moving
part $[A_2]^*$ by the rotation angle $2\pi\times 1/3$.

In the untwisted sector, by definition, the permutation acts as
\begin{align}
\langle p,p',p''|\mbox{Permutation}|p,p',p''\rangle
=\langle p,p',p''||p',p'',p\rangle=\delta_{p,p'}\delta_{p',p''}.
\end{align}
Here we denote the left-moving momenta corresponding to the three $E_6$ parts of 
the even self-dual lattice $(E_6)^3\times[A_2]^*$ as $p, p'$ and $p''$.
Hence, 
only the diagonal $E_6$ remains after the insertion of the permutation operator.
On the other hand, the right-moving sector is removed except for the origin.
As a result, the partition function is
given by $(E+2e)(3\tau)$.
To determine which states survive after the orbifold projection,
it is useful to consider the eigenstates of the permutation
\begin{align}
|p,p',p''\rangle_k=\frac{1}{\sqrt{3}}
\bigl(|p,p',p''\rangle
+\omega^k|p'',p,p'\rangle
+\omega^{2k}|p',p'',p\rangle\bigr),
\label{EigenAdjointStates}
\end{align}
with the eigenvalues $\omega^k$ ($k=0,1,2$).
In the above partition function, there has been cancellation among different states,
\begin{align}
(E+2e)(3\tau)
=(E+2e)(3\tau)+\frac{1+\omega^\beta+\omega^{2\beta}}{3}
\bigl((E^3+2e^3)(\tau)-(E+2e)(3\tau)\bigr),
\end{align}
where the first term on the right-hand side is the contribution of the 
diagonal states $|p,p,p\rangle$, while the second term originates from the other 
eigenstates $|p,p',p''\rangle_k$ with $k=0,1$ and $2$, respectively.
Note that the eigenstates that survive after the orbifold projection depend 
on the phase originating from other factors.
For example, $k\neq0$ states might survive if they are combined with
right-moving states with suitable orbifold phases that cancel the phase $\omega^k$.
This is how the adjoint Higgs field can appear in the untwisted sector
when we increase the Kac-Moody level by performing a permutation,
as discussed generally in \cite{adjoint}. 

Then, we define the partition function in each sector, taking the extra 
phase factors into account, as
\begin{align}
\left[\begin{matrix}
&\displaystyle
\frac{(-i)^3}{9\sqrt{3}}E\Bigl(\frac{\tau}{3}\Bigr)
&\displaystyle
\frac{i^3}{9\sqrt{3}}E\Bigl(\frac{\tau}{3}\Bigr)\\[10pt]
(E+2e)(3\tau)
&\displaystyle
\frac{(-i)^3}{9\sqrt{3}}E\Bigl(\frac{\tau+1}{3}\Bigr)
&\displaystyle
\frac{i^3}{9\sqrt{3}}E\Bigl(\frac{\tau+2}{3}\Bigr)\\[10pt]
(-1)^3(E+2e)(3\tau)
&\displaystyle
\frac{(-i)^3}{9\sqrt{3}}E\Bigl(\frac{\tau+2}{3}\Bigr)
&\displaystyle
\frac{i^3}{9\sqrt{3}}E\Bigl(\frac{\tau+1}{3}\Bigr)
\end{matrix}\right].
\label{withoutshift}
\end{align}
Here note that the arguments of the twisted sector partition functions
are divided by $3$.
Typically, the original arguments contribute significantly to the mass and it
is difficult to obtain massless states in a large representation such as
the adjoint representation in the twisted sector.
Here we increase the Kac-Moody level by performing the permutation
orbifold action for the three $E_6$ lattices, and as a result we obtain a
smaller argument of $\tau/3$.
Even though a larger zero-point energy appears after the orbifold
projection, as a whole, a smaller argument is still preferred in this
case.
This is another way to obtain adjoint Higgs fields by increasing the
Kac-Moody level.

The result, however, contains only the functions $E$ in the twisted
sectors.
Since the matter/antimatter resides in the
fundamental/antifundamental representation, we instead require the functions
$e$, whose lattice is shifted from that of $E$ by a fundamental weight.
As we have seen in the previous subsection, to introduce a
momentum shift in the twisted sector, we need to assign phases to
various states in the untwisted sector by introducing a shift action 
in the coordinate space.
Therefore, we introduce an additional shift, 
$(\bm\omega,\bm\omega,\bm\omega)/3$, with $\bm\omega$ being
the weight vector corresponding to the generator of the conjugacy
classes of $E_6$, so that a momentum shift by the fundamental weight is
realized\footnote{
Precisely speaking, the remaining $E_6$ is slightly different from
the simple diagonal $E_6$ that remains when the additional shift is
not introduced, as discussed in subsection \ref{latphase}. 
Note that this shift action is the only one compatible with the 
diagonal $E_6$ symmetry.}.
Namely, we obtain the partition function 
$(E+\omega e+\omega^2e)(3\tau)=(E-e)(3\tau)$ in the untwisted sector
instead of
$(E+e+e)(3\tau)=(E+2e)(3\tau)$ and introduce the functions $e$ in the
twisted sectors.

Hence, instead of \eqref{withoutshift}, we define the
orbifold partition function for generic $(\alpha,\beta)$ sectors with
${\rm GCD}(\alpha,\beta)\ne 0\mbox{ mod }3$ as
\begin{align}
E\Bigl[\begin{matrix}\alpha\\\beta\end{matrix}\Bigr](\tau)
=\varphi_E
\Bigl[\begin{matrix}(\alpha)_3\\(\beta)_3\end{matrix}\Bigr]
\widehat E
\Bigl[\begin{matrix}\alpha\\\beta\end{matrix}\Bigr](\tau),
\label{Eseparate}
\end{align}
with $(n)_3$ being $n$ mod 3, whose value is $0,1$ or $2$.
As before, we have divided the modular covariant partition function
into the product of the numerical factor $\varphi_E$ and the physical
partition function $\widehat E$.
Here the numerical factor $\varphi_E$ is given by
\begin{align}
\varphi_E
\Bigl[\begin{matrix}(\alpha)_3\\(\beta)_3\end{matrix}\Bigr]
=\left[\begin{matrix}
&\displaystyle\frac{(-i)^3}{9\sqrt{3}}
&\displaystyle\frac{i^3}{9\sqrt{3}}\\[10pt]
1^3&\displaystyle\frac{(-i)^3}{9\sqrt{3}}
&\displaystyle\frac{i^3}{9\sqrt{3}}\\[10pt]
(-1)^3&\displaystyle\frac{(-i)^3}{9\sqrt{3}}
&\displaystyle\frac{i^3}{9\sqrt{3}}
\end{matrix}\right],
\end{align}
while the physical partition function $\widehat E$ is given by
\begin{align}
\widehat E\Bigl[\begin{matrix}\alpha\\
\beta\end{matrix}\Bigr](\tau)
=\left[\begin{matrix}
&\displaystyle e\Bigl(\frac{\tau}{3}\Bigr)
&\displaystyle e\Bigl(\frac{\tau}{3}\Bigr)\\[10pt]
(E-e)(3\tau)
&\displaystyle e\Bigl(\frac{\tau+1}{3}\Bigr)
&\displaystyle\omega^2 e\Bigl(\frac{\tau+2}{3}\Bigr)\\[10pt]
(E-e)(3\tau)
&\displaystyle e\Bigl(\frac{\tau+2}{3}\Bigr)
&\displaystyle e\Bigl(\frac{\tau+1}{3}\Bigr)
\end{matrix}\right],
\label{physEmod3}
\end{align}
for $\alpha,\beta=0,1,2$.
We further define the partition function $\widehat E$ for the other 
values of $(\alpha,\beta)$ as
\begin{align}
\widehat E\Bigl[\begin{matrix}\alpha\\\beta\end{matrix}\Bigr](\tau)
=\omega^{-\alpha[\beta/3]+\beta[\alpha/3]}
\widehat E\Bigl[\begin{matrix}(\alpha)_3\\
(\beta)_3\end{matrix}\Bigr](\tau).
\label{physE}
\end{align}
Here $[x]$ is given by the largest integer that does not
exceed $x$.
Note that, although the same function $e(\tau/3)$ appears in
both the $\alpha=1$ and $\alpha=2$ sectors, these functions actually correspond to
the root lattices shifted differently by one and two fundamental
weights, respectively.

The extra factor $\omega^2$ in the $(2,1)$ sector in \eqref{physEmod3}
may seem new.
However, it appears naturally by considering that the states in the
$(2,1)$ sector acquire half the phases of those in the $(2,2)$ sector,
which is obtained directly from the ${\mathcal T}$-transformation of
the $(2,0)$ sector.
As a result, the total partition function $E\Bigl[\begin{matrix}\alpha\\\beta\end{matrix}\Bigr](\tau)$ satisfies the elegant modular transformation
property
\begin{align}
E\Bigl[\begin{matrix}\alpha\\\beta\end{matrix}\Bigr](\tau+1)
=E\Bigl[\begin{matrix}\alpha\\\beta+\alpha\end{matrix}\Bigr](\tau),
\quad
E\Bigl[\begin{matrix}\alpha\\\beta\end{matrix}\Bigr](-1/\tau)
=\sqrt{-i\tau}^6i^3
E\Bigl[\begin{matrix}\beta\\-\alpha\end{matrix}\Bigr](\tau),
\end{align}
as before.
Compared with \eqref{noshift}, the extra power of $3$ of $i$ in the
${\mathcal S}$-transformation originates from the three complex dimensions of
the diagonal $E_6$ lattice.
In the ${\mathbb Z}_N$ $(N\in 3{\mathbb Z})$ orbifold, the periodic
relation is given by 
\begin{align}
E\Bigl[\begin{matrix}\alpha\\\beta+N\end{matrix}\Bigr](\tau)
=e^{\frac{4}{9}\pi i N\alpha}
E\Bigl[\begin{matrix}\alpha\\\beta\end{matrix}\Bigr](\tau).
\label{periodE6}
\end{align}

To summarize, in this subsection we have shown that we
can obtain an adjoint Higgs field by increasing the Kac-Moody level,
which is achieved by the orbifold action permuting
three $E_6$ lattices.
Furthermore, to obtain a nonvanishing
generation number we need to introduce shifts to break the symmetry
between chiral and antichiral matter.

\section{Models and partition functions}
As we explained in the introduction, in the asymmetric orbifold
construction it is necessary to compactify our heterotic string
theory on a general Lorentzian (22,6)-dimensional even self-dual
lattice and consider an orbifold action that can act on
left-moving and right-moving modes independently.
As reviewed in the introduction, in \cite{letter} we chose the even self-dual
lattice $[(A_2)^2\times(E_6)^3]\times[E_6]^*$ to compactify our
heterotic string theory and considered a ${\mathbb Z}_{12}$ orbifold
action that acts on each factor of the lattice independently as
follows:
\begin{itemize}
\item $(A_2)^2$: an arbitrary ${\mathbb Z}_{12}$ shift action (which
will be constrained later) or the twist action with the rotation angle
$2\pi\times 1/3$ (or the $1/3$-twist hereafter)\footnote{
Although we have also studied the possibility of the Weyl reflection,
we found that it does not lead to modular invariant
partition functions. Thus, we do not consider this possibility in the following.}
\item $(E_6)^3$: the ${\mathbb Z}_3$ permutation action with the shift 
\item $[E_6]^*$: the ${\mathbb Z}_{12}$ Coxeter element of $E_6$,
namely, the twist action with the rotation angles
\begin{align}
2\pi\times(1/12,-5/12,1/3).
\label{E6coxeter}
\end{align}
\end{itemize}
In this section, we study their partition functions explicitly.

\subsection{Unorbifolded theory}\label{sec:Unorbifolded theory}
The even self-dual lattice $[(A_2)^2\times(E_6)^3]\times[E_6]^*$ can be 
generated from the $E_8$ lattice as explained in \eqref{lattice}.
The $E_8$ lattice is represented by the conjugacy classes 
$\Pi_{E_8}=\oplus_{k=0}^2k(1,1)$ 
in $E_6\times A_2$.
Following the construction, we find that the conjugacy classes
in $E_6\times A_2$ (and also in $E_6\times[E_6]^*$) are
changed into the conjugacy classes
\begin{align}
\Pi=
\oplus_{k,l=0}^2\bigl[l(1,1,1,0)+k(0,1,2,1)\bigr] \label{invset}
\end{align}
in $(A_2)^3\times[E_6]^*$ and
$(A_2)^2\times[E_6\times E_6]^*$ and subsequently become
\begin{align}
\oplus_{k,l,m=0}^2
\bigl[m(0,0,1,1,1,0)+l(1,1,0,1,2,0)+k(0,1,0,2,1,1)\bigr]
\label{OriginalLattice}
\end{align}
in $(A_2)^2\times[(A_2)^3\times E_6]^*$ and
$[(A_2)^2\times(E_6)^3]\times[E_6]^*$.
{}From this knowledge, we see that the partition function before orbifolding
is given by
\begin{align}
\Theta^{\rm L}(\tau)
=\sum_{(k_{A1},k_{A2},k_{\mathcal E},k_E)\in\Pi}
A_{k_{A1}}(\tau)
A_{k_{A2}}(\tau)
{\mathcal E}_{k_{\mathcal E}}(\tau)
[E_{k_E}(\tau)]^*.
\label{unorbifold}
\end{align}
Here we have used the conjugacy classes $\Pi$ in the notation of
$(A_2)^3\times[E_6]^*$ before converting one of the $A_2$ into
$(E_6)^3$, and
we have defined ${\mathcal E}_k$, the partition functions for the $(E_6)^3$ part, as
\begin{align}
\bigl({\mathcal E}\bigr)_k=\bigl(E^3+2e^3,3Ee^2,3Ee^2\bigr).
\label{calE}
\end{align}
The integers $k=0, 1$ and $2$ of ${\mathcal E}_k$ correspond to the conjugacy classes of
$A_2$ which can be translated to the conjugacy classes of $(E_6)^3$
by the lattice engineering technique, as
$A_2\to[E_6]^*\to[(A_2)^3]^*\to(E_6)^3$.
The modular transformation of ${\mathcal E}_k$ is given by
\begin{align}
{\mathcal E}_k(\tau+1)=\omega^{k^2}{\mathcal E}_k(\tau),\quad
{\mathcal E}_k(-1/\tau)=\sqrt{-i\tau}^{18}
\frac{1}{\sqrt{3}}\sum_{l=0}^2\omega^{-kl}{\mathcal E}_l(\tau).
\label{modularcalEe}
\end{align}
Namely, as in \eqref{AEmodular}, under the 
${\mathcal T}$-transformation the partition function ${\mathcal E}_k$
acquires a factor $\omega^{k^2}$, while the
${\mathcal S}$-transformation takes the form of a finite Fourier
transformation.
The full partition function with the lattice part given by
\eqref{unorbifold} is clearly modular invariant up to the power of 
$\sqrt{-i\tau}$ from the construction of this lattice.

\subsection{${\mathbb Z}_4$ suborbifold}
Next, we consider the partition functions for orbifold theory.
Their form strongly depends on whether $\beta=0\mbox{ mod 3}$ or
$\beta\ne 0\mbox{ mod 3}$,
because some components of the twist
actions become trivial for
$\beta=0\mbox{ mod 3}$.
Therefore, we examine the two cases separately.
In this subsection, we concentrate on the ${\mathbb Z}_4$ suborbifold, namely,
the case with $g=\mbox{GCD}(\alpha,\beta)=0\mbox{ mod 3}$.
In the following, we first discuss the untwisted sectors and then 
define the partition functions for all of the
$(\alpha,\beta)$-sectors, so that each sector of the partition
functions is modular covariant.

First, we consider the effect of the twist part of the orbifold action.
In the untwisted sector, the last component of the twist
\eqref{E6coxeter} in $[E_6]^*$ becomes trivial and keeps the last factor intact.
This factor corresponds to the $\widetilde A_2$ plane of the decomposition 
$E_6\to D_4\times\widetilde A_2$,
described in appendix \ref{lattice_decompD4A2}, while the first
two rotations act on the $D_4$ planes and extract the origin.
In the lattice decomposition \eqref{D4A2}, 
we find that the conjugacy classes of $\widetilde A_2$ paired with the
origin of the $D_4$ plane are 
exactly the same as the corresponding classes of $E_6$.
Therefore, the surviving partition function is that of
$[(A_2)^2\times(E_6)^3]\times[{\widetilde A_2}]^*$ summed over the
same conjugacy classes as $\Pi$ in \eqref{invset}.
Note that in the two $A_2$ lattices we include general shifts to
allow more possibilities in model construction, since the inclusion does
not change the modular transformation as we have seen in \eqref{A2shift}.

Here, we note a beautiful property
of the partition function under the modular transformation.
We assume that the partition function consists of four triplet
building blocks,
\begin{align}
\Theta(\tau)=\sum_{(k_A,k_B,k_C,k_D)\in\Pi}
A_{k_A}(\tau)B_{k_B}(\tau)C_{k_C}(\tau)D_{k_D}(\tau).
\end{align}
Here $k_M\;(M=A,B,C,D)$ runs over $0,1,2$.
If each building block transforms as
\begin{align}
M_{k_M}(\tau+1)=\omega^{k_M^2}M_{k_M}(\tau)
\label{omegak2}
\end{align}
under the ${\mathcal T}$-transformation, the partition function is
obviously invariant.
Suppose under the ${\mathcal S}$-transformation, each building block
transforms as a finite Fourier transformation:
\begin{align}
M_{k_M}(-1/\tau)
=\frac{1}{\sqrt{3}}\sum_{l_M=0}^2\omega^{-k_Ml_M}\widetilde M_{l_M}(\tau),
\label{Fourier}
\end{align}
where we have omitted the power of the factor $\sqrt{-i\tau}$
assuming that it will be canceled out finally.
Then, we find that the partition function transforms into
\begin{align}
\Theta(-1/\tau)
=\sum_{(l_A,l_B,l_C,l_D)\in\Pi}
\widetilde A_{l_A}(\tau)\widetilde B_{l_B}(\tau)
\widetilde C_{l_C}(\tau)\widetilde D_{l_D}(\tau).
\end{align}
Namely, under the ${\mathcal S}$-transformation each building block $M$ 
in the partition function effectively transforms into $\widetilde M$. 
In particular, if $M=\widetilde M$ for each building block, the partition 
function is modular invariant, which is the case of \eqref{unorbifold}.
This modular invariance is not a surprise but simply a consequence of
the two subsequent lattice decompositions $E_8\to E_6\times A_2$ and 
$E_6\to(A_2)^3$.

We now define the partition function of the ${\mathbb Z}_4$
suborbifold as
\begin{align}
\Theta^{\rm L}\Bigl[\begin{matrix}\alpha\\\beta\end{matrix}\Bigr](\tau)
=\sum_{(k_{A1},k_{A2},k_{\mathcal E},k_{\mathcal A})\in\Pi}
A_{k_{A1}}^{\vec l_1}
\Bigl[\begin{matrix}\alpha\\\beta\end{matrix}\Bigr]
(\tau)
A_{k_{A2}}^{\vec l_2}
\Bigl[\begin{matrix}\alpha\\\beta\end{matrix}\Bigr]
(\tau)
{\mathcal E}_{k_{\mathcal E}}
(\tau)
\Bigl[{\mathcal A}_{k_{\mathcal A}}
\Bigl[\begin{matrix}\alpha\\\beta\end{matrix}\Bigr]
(\tau)\Bigr]^*.
\label{Zfourtriplets}
\end{align}
Compared with the partition function in
unorbifolded theory \eqref{unorbifold}, which is 
the (0,0) sector of orbifold theory,
the first and second entries
$A_{k}$ are replaced by $A_{k}^{\vec l}$ \eqref{PhysAdef} with the
shift contribution, 
where $\vec l$ is a general vector representing the ${\mathbb Z}_{12}$ shift 
determined later in \eqref{ShiftVector}, 
while the third entry ${\mathcal E}_{k}$ \eqref{calE}
is exactly the same as that in \eqref{unorbifold} regardless of the
$(\alpha,\beta)$ sector.
Note that, under the ${\mathcal T}$-transformation, these factors
acquire the phases $\omega^{k^2}$ as in \eqref{omegak2}, while their
${\mathcal S}$-transformations take the form of the finite Fourier
transformation \eqref{Fourier}.
The final entry in \eqref{Zfourtriplets}, 
$\left[{\mathcal A}_{k_{\mathcal A}}\right]^*$, 
is the contribution from $[{\widetilde A_2}]^*$.
Since it originates from the right-moving part,  
we have taken the complex conjugate in \eqref{Zfourtriplets}.
${\mathcal A}_k$ is defined as follows\footnote{
The components in the matrix on the right-hand side should be
chosen as $(\alpha/3)_4$ and $(\beta/3)_4$.}:
\begin{align}
{\mathcal A}_k\Bigl[\begin{matrix}\alpha\\\beta\end{matrix}\Bigr]
=\left[\begin{matrix}
&\bigl({\mathcal A}^1_0\bigr)_k&\bigl({\mathcal A}^1_0\bigr)_k
&\bigl({\mathcal A}^1_0\bigr)_k\\
 \bigl({\mathcal A}^0_1\bigr)_k&\bigl({\mathcal A}^1_1\bigr)_k
&\bigl({\mathcal A}^0_1\bigr)_k&\bigl({\mathcal A}^1_1\bigr)_k\\
 \bigl({\mathcal A}^0_1\bigr)_k&\bigl({\mathcal A}^1_0\bigr)_k
&\bigl({\mathcal A}^1_1\bigr)_k&\bigl({\mathcal A}^1_0\bigr)_k\\
 \bigl({\mathcal A}^0_1\bigr)_k&\bigl({\mathcal A}^1_1\bigr)_k
&\bigl({\mathcal A}^0_1\bigr)_k&\bigl({\mathcal A}^1_1\bigr)_k
\end{matrix}\right],
\label{tildeA}
\end{align}
with components $({\mathcal A}^\bullet_\bullet)_k$ defined by
\begin{align}
\bigl({\mathcal A}^0_1\bigr)_k
&=\bigl(\widehat{\mathcal A}^0_1\bigr)_k,&
\bigl(\widehat{\mathcal A}^0_1\bigr)_k
&=\Bigl(A\bigl(2\tau\bigr),
a\bigl(2\tau\bigr),a\bigl(2\tau\bigr)\Bigr),
\nonumber\\
\bigl({\mathcal A}^1_0\bigr)_k
&=-\frac{1}{2}\bigl(\widehat{\mathcal A}^1_0\bigr)_k,&
\bigl(\widehat{\mathcal A}^1_0\bigr)_k
&=\Bigl(A\bigl(\tau/2\bigr),
a\bigl(\tau/2\bigr),
a\bigl(\tau/2\bigr)\Bigr),
\nonumber\\
\bigl({\mathcal A}^1_1\bigr)_k
&=-\frac{1}{2}\bigl(\widehat{\mathcal A}^1_1\bigr)_k,&
\bigl(\widehat{\mathcal A}^1_1\bigr)_k
&=\Bigl(A\bigl((\tau+1)/2\bigr),
\omega a\bigl((\tau+1)/2\bigr),
\omega a\bigl((\tau+1)/2\bigr)\Bigr).
\label{tildeAdef}
\end{align}
In other words, we have separated the partition functions ${\mathcal A}$
into the numerical factors $\varphi_{\mathcal A}$ and the
physical partition functions $\widehat{\mathcal A}$ as
\begin{align}
{\mathcal A}_k
\Bigl[\begin{matrix}\alpha\\\beta\end{matrix}\Bigr]
=\varphi_{\mathcal A}
\Bigl[\begin{matrix}(\alpha)_{12}\\(\beta)_{12}\end{matrix}\Bigr]
\widehat{\mathcal A}_k
\Bigl[\begin{matrix}\alpha\\\beta\end{matrix}\Bigr],
\label{calAseparate}
\end{align}
\begin{align}
\varphi_{\mathcal A}
\Bigl[\begin{matrix}(\alpha)_{12}\\(\beta)_{12}\end{matrix}\Bigr]
=\left[\begin{matrix}
1&-1/2&-1/2&-1/2\\
1&-1/2&1&-1/2\\
1&-1/2&-1/2&-1/2\\
1&-1/2&1&-1/2
\end{matrix}\right],\quad
\widehat{\mathcal A}_k
\Bigl[\begin{matrix}\alpha\\\beta\end{matrix}\Bigr]
=\left[\begin{matrix}
&\bigl(\widehat{\mathcal A}^1_0\bigr)_k
&\bigl(\widehat{\mathcal A}^1_0\bigr)_k
&\bigl(\widehat{\mathcal A}^1_0\bigr)_k\\
 \bigl(\widehat{\mathcal A}^0_1\bigr)_k
&\bigl(\widehat{\mathcal A}^1_1\bigr)_k
&\bigl(\widehat{\mathcal A}^0_1\bigr)_k
&\bigl(\widehat{\mathcal A}^1_1\bigr)_k\\
 \bigl(\widehat{\mathcal A}^0_1\bigr)_k
&\bigl(\widehat{\mathcal A}^1_0\bigr)_k
&\bigl(\widehat{\mathcal A}^1_1\bigr)_k
&\bigl(\widehat{\mathcal A}^1_0\bigr)_k\\
 \bigl(\widehat{\mathcal A}^0_1\bigr)_k
&\bigl(\widehat{\mathcal A}^1_1\bigr)_k
&\bigl(\widehat{\mathcal A}^0_1\bigr)_k
&\bigl(\widehat{\mathcal A}^1_1\bigr)_k
\end{matrix}\right].
\end{align}
Thanks to the elegant modular transformation properties of the
quantities $({\mathcal A}^\bullet_\bullet)_k$,
\begin{align}
{\mathcal T}&:\bigl({\mathcal A}^0_1\bigr)_k
\mapsto\omega^{-k^2}\bigl({\mathcal A}^0_1\bigr)_k,&
{\mathcal S}&:\bigl({\mathcal A}^0_1\bigr)_k
\mapsto-\sqrt{-i\tau}^2\sum_{l=0}^2\frac{\omega^{kl}}{\sqrt{3}}
\bigl({\mathcal A}^1_0\bigr)_l,\nonumber\\
{\mathcal T}&:\bigl({\mathcal A}^1_0\bigr)_k
\mapsto\omega^{-k^2}\bigl({\mathcal A}^1_1\bigr)_k,&
{\mathcal S}&:\bigl({\mathcal A}^1_0\bigr)_k
\mapsto-\sqrt{-i\tau}^2\sum_{l=0}^2\frac{\omega^{kl}}{\sqrt{3}}
\bigl({\mathcal A}^0_1\bigr)_l,\nonumber\\
{\mathcal T}&:\bigl({\mathcal A}^1_1\bigr)_k
\mapsto\omega^{-k^2}\bigl({\mathcal A}^1_0\bigr)_k,&
{\mathcal S}&:\bigl({\mathcal A}^1_1\bigr)_k
\mapsto-\sqrt{-i\tau}^2\sum_{l=0}^2\frac{\omega^{kl}}{\sqrt{3}}
\bigl({\mathcal A}^1_1\bigr)_l,
\label{tildeAmodular}
\end{align}
$[{\mathcal A}]^*$ also transforms as in \eqref{omegak2} and 
\eqref{Fourier} except for the minus sign originating
from the right-hand side of \eqref{tildeAmodular}.

We find that the total lattice partition function
\eqref{Zfourtriplets} is covariant under the modular transformation
up to the above minus sign.
As we have explained in appendix \ref{osc_pf}, the twisted boson partition function
acquires an extra factor $i$ in the ${\mathcal S}$-transformation for
each complex dimension.
To make the partition function transform covariantly, we have
required the lattice partition function to transform in the same way.
Here the surviving lattice
$[(A_2)^2\times(E_6)^3]\times[{\widetilde A_2}]^*$ 
spans $(2\times2+3\times6)/2=11$ complex dimensions with two $A_2$
lattices and three $E_6$ lattices in the left-moving part and one
complex dimension with one $\widetilde A_2$ lattice in the
right-moving part.
Therefore, the minus sign acquired in the
${\mathcal S}$-transformation \eqref{tildeAmodular} matches the
required phase $i^{11}(i^1)^*=-1$ effectively.

Combined with the fermion and twisted boson partition functions,
$\Theta^{\rm F}$ and $\Theta^{\rm B}$, respectively,
the total partition function in the ${\mathbb Z}_4$
suborbifold is given by
\begin{align}
&Z\Bigl[\begin{matrix}\alpha\\\beta\end{matrix}\Bigr]
=
\frac{1}{{\rm Im}\,\tau}
\varphi
\Bigl[\begin{matrix}(\alpha)_{12}\\(\beta)_{12}\end{matrix}\Bigr]
\cdot\frac{1}{\eta^{24}}
\cdot\Theta^{\rm L}
\Bigl[\begin{matrix}\alpha\\\beta\end{matrix}\Bigr]
\cdot\biggl[\frac{1}{\eta^4}
\Theta^{\rm B}_{(1/12)}
\Bigl[\begin{matrix}\alpha\\\beta\end{matrix}\Bigr]
\Theta^{\rm B}_{(-5/12)}
\Bigl[\begin{matrix}\alpha\\\beta\end{matrix}\Bigr]
\Theta^{\rm F}_{(0,1/12,-5/12,1/3)}
\Bigl[\begin{matrix}\alpha\\\beta\end{matrix}\Bigr]
\biggr]^*.
\label{Z4TotalPF}
\end{align}
Here, the first factor, $1/{\rm Im}\, \tau$, originates from the integration 
over the transverse momenta in the four-dimensional spacetime, and the 
definition of the overall prefactor $\varphi$ will be given later 
in this section.
Finally, we have to ensure the orbifold periodic condition 
(\ref{orbperiod}).
Since the periodic relation of the lattice partition function is given 
by \eqref{periodA2} or
\eqref{periodE6}, to satisfy the orbifold periodic condition,
we require
\begin{align}
\frac{N(=12)\times3}{2}\biggl[\sum_{j=1}^2\vec l_j\cdot C\vec l_j\biggr]\in{\mathbb Z},
\label{z4_periodic}
\end{align}
where the factor $3$ in the numerator is due to the fact that 
$\alpha$ is now a multiple of $3$.
Note that the contributions from the right-moving part cancel among
themselves as we have noted below \eqref{bosonperiod} in appendix \ref{osc_pf}.

\subsection{${\mathbb Z}_{12}$ orbifold}
We now proceed to the other sectors of the ${\mathbb Z}_{12}$ orbifold,
namely, the
$(\alpha,\beta)$ sectors with
${\rm GCD}(\alpha,\beta)\ne 0\mbox{ mod }3$.
These sectors can be generated from the untwisted sector with
$\beta\ne 0\mbox{ mod }3$.
In the untwisted sector,
the orbifold action twisting the right-moving $[E_6]^*$
projects out all the states except for the origin, and the orbifold action
permuting $(E_6)^3$ extracts the diagonal contribution
\begin{align}
\oplus_{m=0}^2m(0,0,1^{\rm diag}),
\label{Z3InvariantSubLattice}
\end{align}
where the components denote the conjugacy classes of the two $A_2$
lattices and the diagonal $E_6$ lattice.
Therefore, the partition function contains only the $(A_2)^2$ root
lattice with shifts and the diagonal $(E_6)^3$ lattice $(E-e)(3\tau)$.
Although the setups used for the orbifold construction in subsections
\ref{shiftaction} and \ref{permutation} are different, the resulting
conjugacy classes of the two $A_2$ lattices and the diagonal $E_6$
lattice are exactly the same, and therefore we can utilize the partition
function studied there.
If we choose the $1/3$-twist action for the $A_2$ part instead of 
the shift actions, the contribution must be replaced with the twisted boson
partition function.

We first consider the shift actions as the orbifold action on the
$(A_2)^2$ part and define the partition functions as
\begin{align}
\Theta^{\rm L}\Bigl[\begin{matrix}\alpha\\\beta\end{matrix}\Bigr](\tau)
=\Bigl(
\prod_{j=1}^2
A^{\vec l_j}\Bigl[\begin{matrix}\alpha\\\beta\end{matrix}\Bigr](\tau)
\Bigr)
E\Bigl[\begin{matrix}\alpha\\\beta\end{matrix}\Bigr](\tau).
\label{Z12LatticePF}
\end{align}
The definition of each building block has already appeared in
\eqref{shift} and \eqref{Eseparate}.
As we have seen there, each building block satisfies an elegant modular
transformation property.
To summarize, our partition function takes the form
\begin{align}
&Z\Bigl[\begin{matrix}\alpha\\\beta\end{matrix}\Bigr]
=
\frac{1}{{\rm Im}\,\tau}
\varphi
\Bigl[\begin{matrix}(\alpha)_{12}\\(\beta)_{12}\end{matrix}\Bigr]
\cdot\frac{1}{\eta^{12}}
\Bigl(\Theta^{\rm B}_{(2/3)}
\Bigl[\begin{matrix}\alpha\\\beta\end{matrix}\Bigr]\Bigr)^6
\cdot\Theta^{\rm L}
\Bigl[\begin{matrix}\alpha\\\beta\end{matrix}\Bigr]\nonumber\\
&\quad
\cdot\biggl[\frac{1}{\eta^2}
\Theta^{\rm B}_{(1/12)}
\Bigl[\begin{matrix}\alpha\\\beta\end{matrix}\Bigr]
\Theta^{\rm B}_{(-5/12)}
\Bigl[\begin{matrix}\alpha\\\beta\end{matrix}\Bigr]
\Theta^{\rm B}_{(1/3)}
\Bigl[\begin{matrix}\alpha\\\beta\end{matrix}\Bigr]
\Theta^{\rm F}_{(0,1/12,-5/12,1/3)}
\Bigl[\begin{matrix}\alpha\\\beta\end{matrix}\Bigr]
\biggr]^*.
\label{Z12TotalPF}
\end{align}
In this case, the orbifold periodic condition is given by
\begin{align}
\frac{N(=12)}{2}\biggl[\sum_{j=1}^2\vec l_j\cdot C\vec l_j
+4/9
\biggr]\in{\mathbb Z},
\label{periodic}
\end{align}
where we have omitted the contribution from
the left-moving twisted bosons \eqref{bosonperiod}, which becomes
integral in this case.
Note that the periodic condition for ${\mathbb Z}_4$ suborbifold \eqref{z4_periodic}
is automatically satisfied provided the above condition is satisfied.

For the cases that the orbifold action on the $A_2$ part is the
$1/3$-twist, we should replace the factor $A^{\vec l}/\eta^2$ in the
above partition function with $\Theta^{\rm B}_{(1/3)}$, which
contributes to the periodic condition \eqref{periodic} as
$-N(1/3)^2/2$ instead of $N\vec l\cdot C\vec l/2$.

\subsection{Prefactor}\label{prefactor}
In this subsection, we discuss the prefactor $\varphi$, which has been postponed so far.
For the case of the shift action
without any twist on $(A_2)^2$,  we define it as
\begin{align}
\varphi
\Bigl[\begin{matrix}(\alpha)_{12}\\(\beta)_{12}\end{matrix}\Bigr]
=\begin{cases}-4&{\rm if}\;g=6\\
-81\sqrt{3}&{\rm if}\;g=4\\
2&{\rm if}\;g=3\\
27\sqrt{3}&{\rm if}\;g=2\\
-27\sqrt{3}&{\rm if}\;g=1
\end{cases},
\end{align}
with $g={\rm GCD}(\alpha,\beta,12)$.
Note that since the partition function in each $(\alpha,\beta)$ sector
is modular covariant, the prefactor $\varphi$ has to be common in the
$(\alpha,\beta)$ sectors that share the same value of $g$ and hence
are related by a modular transformation.
The prefactor is determined so that it cancels the unwanted numerical
factors such as $2\sin\beta\pi\phi$ in the untwisted sectors:
\begin{align}
&\bigl([2\sin\beta\pi(2/3)]^6\cdot
2\sin\beta\pi(1/12)\cdot
2\sin\beta\pi(-5/12)\cdot
2\sin\beta\pi(1/3)
\bigr)_{\beta=0,1,\cdots,11}\nonumber\\
&\quad=(0,-27\sqrt{3},-27\sqrt{3},0,-81\sqrt{3},27\sqrt{3},
0,-27\sqrt{3},81\sqrt{3},0,27\sqrt{3},27\sqrt{3}),\\
&\bigl(2\sin\beta\pi(1/12)\cdot
2\sin\beta\pi(-5/12)
\bigr)_{\beta=0,3,6,9}=(0,2,-4,2).
\end{align}
Since we have three lattice partition functions with extra minus signs
in the untwisted sectors with $\beta=2\mbox{ mod }3$, we have to
change signs for these cases to obtain the prefactor $\varphi$.

In the case with twists, we should multiply the above result by
$[2\sin\beta\pi(1/3)]^{1\,{\rm or}\,2}$ and the extra minus signs
appropriately.
The prefactors for the cases with one twist and two twists on $(A_2)^2$
are given respectively by
\begin{align}
\varphi
\Bigl[\begin{matrix}(\alpha)_{12}\\(\beta)_{12}\end{matrix}\Bigr]
=\begin{cases}-4&{\rm if}\;g=6\\
243&{\rm if}\;g=4\\
2&{\rm if}\;g=3\\
-81&{\rm if}\;g=2\\
-81&{\rm if}\;g=1
\end{cases},
\quad
\varphi
\Bigl[\begin{matrix}(\alpha)_{12}\\(\beta)_{12}\end{matrix}\Bigr]
=\begin{cases}-4&{\rm if}\;g=6\\
-243\sqrt{3}&{\rm if}\;g=4\\
2&{\rm if}\;g=3\\
81\sqrt{3}&{\rm if}\;g=2\\
-81\sqrt{3}&{\rm if}\;g=1
\end{cases}.
\end{align}

\subsection{Classification}
In the previous subsections, we defined modular
invariant partition functions for our models.
Neither the extra shifts $\vec l_i$ nor the $1/3$-twist in the
$(A_2)^2$ lattices changes the modular invariance provided the
periodic condition \eqref{periodic} (or the condition with the replacement of
$N\vec l\cdot C\vec l/2$ with $-N(1/3)^2/2$) is satisfied.
In this subsection, we investigate the number of consistent
models.

Regarding the shift action, since we are considering the
${\mathbb Z}_{12}$ orbifold, we can choose the shift vector to be
$\vec l=(m^1,m^2)/12$ in each of the two $A_2$ planes with
$m^1,m^2=0,1,\cdots,11$, because 12 times the shift belongs to the 
root lattice and the shift is defined up to the root lattice.
In the following we define the shift vectors in the lattice
space as
\begin{align}
{\bm\lambda}=\vec{\bm\alpha}\cdot\vec l
=\frac{{\bm\alpha}_1m^1+{\bm\alpha}_2m^2}{12}
\end{align}
for convenience.
Noting that the possible values of
\begin{align}
L=\frac{1}{2}12^2\bm\lambda\circ\bm\lambda
=\frac{1}{2}12^2\vec l\cdot C\vec l=(m^1)^2-m^1m^2+(m^2)^2
\end{align}
are
\begin{align}
L=0,1,3,4,7,9
\quad\mbox{mod 12},
\end{align}
we have to take the combinations $\{L=4,L=0\}$, $\{L=1,L=3\}$ and
$\{L=7,L=9\}$ in the two $A_2$ planes so that their sum is $4$ mod
$12$ to satisfy the periodic condition \eqref{periodic}.
There are 24 combinations of $(m^1,m^2)$ resulting in $L=4$ while 12
combinations result in $L=0$. 
Similarly, there are $36\times 18$ combinations for the
$\{L=1,L=3\}$ condition and the $\{L=7,L=9\}$ condition.
Thus, there appear to be many possibilities:
$24\times 12+36\times 18+36\times 18=1584$.
However, owing to the root lattice symmetry
$(m^1,m^2)\sim(m^1+12,m^2)$, $(m^1,m^2)\sim(m^1,m^2+12)$, the Weyl
reflection symmetry $W_{\alpha_1}:(m^1,m^2)\mapsto(-m^1,m^1+m^2)$,
$W_{\alpha_2}:(m^1,m^2)\mapsto(m^1+m^2,-m^2)$ and the charge
conjugation ${\mathcal C}:(m^1,m^2)\mapsto(-m^1,-m^2)$,
most of them lead to identical models.
After identifying the shift vectors that give the same physical model,
we find there are $3\times 4+4\times 3+3\times 2=30$
models, where the representative elements for each value
of $L$ are given by 
\begin{itemize}
\item the $\{L=4,L=0\}$ case, \vspace{1mm}\\
\begin{tabular}{l}
 -- for $L=4$ :\,(2,0),\,(6,8),\,(4,0), \vspace{1mm}\\
 -- for $L=0$ :\,(0,0),\,(4,8),\,(6,0),\,(10,8), 
\end{tabular}
\item the $\{L=1,L=3\}$ case, \vspace{1mm}\\
\begin{tabular}{l}
 -- for $L=1$ :\,(1,0),\,(5,8),\,(5,0),\,(9,8), \vspace{1mm}\\
 -- for $L=3$ :\,(3,6),\,(7,2),\,(11,10),
\end{tabular}
\item the $\{L=7,L=9\}$ case, \vspace{1mm}\\
\begin{tabular}{l}
 -- for $L=7$ :\,(1,6),\,(5,2),\,(9,10), \vspace{1mm}\\
 -- for $L=9$ :\,(3,0),\,(7,8). 
\end{tabular}
\end{itemize}
In addition, we can identify two shifts whose difference resides on the weight
lattice, because in sectors with $\alpha=0\mbox{ mod }3$, the
difference vector is three times a weight vector and resides exactly
on the root lattice, while in sectors with $\alpha\ne 0\mbox{ mod }3$,
the right-movers are twisted nontrivially and thus the 
left-movers remain only in the root lattice.
For this reason, we can further identify shifts with a difference in
the fundamental weight of $(4,8)$,
\begin{align}
&(2,0)\sim(6,8),\quad
(0,0)\sim(4,8),\quad
(6,0)\sim(10,8),\nonumber\\
&(1,0)\sim(5,8),\quad
(5,0)\sim(9,8),\quad
(3,6)\sim(7,2)\sim(11,10),\nonumber\\
&(1,6)\sim(5,2)\sim(9,10),\quad
(3,0)\sim(7,8).
\end{align}
Furthermore, a shift by (1,0) and a shift by
(5,0) have the same effect, because both $1$ and $5$ are generators in
${\mathbb Z}_{12}$ and the exchange between them simply corresponds to
the exchange among sectors.

For the twist action, since the $1/3$-twist gives the same contribution to the
periodic condition as the shift action with $L=4$, the combination of
the twist and the shift action with $L=0$ satisfies the periodic
condition.

Finally, we are left with only $3\times 2+1\times 1+1\times 1=8$
models:
\begin{align}
\{(2,0),(4,0),\mbox{``rot''}\}\otimes\{(0,0),(6,0)\},\quad
(1,0)\otimes(3,6),\quad(1,6)\otimes(3,0),
\end{align}
where ``rot'' denotes the $1/3$-twist action.
Out of these eight models, only three of them actually contain three
generations.
Their shifts are given by
\begin{align}
(2,0)\otimes(6,0):&\quad
\bm\lambda_1^{(1)}=\frac{\bm\alpha_1}{6},\quad
\bm\lambda_2^{(1)}=\frac{\bm\alpha_1}{2},\nonumber\\
(1,0)\otimes(3,6):&\quad
\bm\lambda_1^{(2)}=\frac{\bm\alpha_1}{12},\quad
\bm\lambda_2^{(2)}=\frac{\bm\alpha_1+2\bm\alpha_2}{4},\nonumber\\
(1,6)\otimes(3,0):&\quad
\bm\lambda_1^{(3)}=\frac{\bm\alpha_1+6\bm\alpha_2}{12},\quad
\bm\lambda_2^{(3)}=\frac{\bm\alpha_1}{4}.
\label{ShiftVector}
\end{align}
We call these models Models 1, 2 and 3 respectively.
Since Model 1 results in the same massless spectrum as the model found
in \cite{KTE6}, we shall restrict ourselves to Models 2 and 3
hereafter.

\section{Analysis of models}
In the previous section, we completely fixed the model setup, so that it satisfies 
the requirements mentioned in the introduction.
In this section, we identify their massless spectra
by detecting the states whose total
phases cancel.
Note that, in addition to the phases originating from the physical partition
functions, various numerical factors contribute to the phases, 
which may be interpreted as fixed point numbers associated with phases.

\subsection{Phases from fixed points}
In the study of the partition functions, we have separated the modular covariant
partition functions into the products of the physical partition functions
and the extra factors as in \eqref{shift}, \eqref{Eseparate},
\eqref{calAseparate} and \eqref{bosonseparate}.
Here we collect all of these extra factors and interpret them as fixed points with phases, although the
geometric picture of fixed points is not very clear in the asymmetric
orbifold.

For this purpose, we first collect the extra numerical factors
from the lattices and twisted bosons as
\begin{align}
\varphi^{\rm L}
\Bigl[\begin{matrix}(\alpha)_{12}\\(\beta)_{12}\end{matrix}\Bigr]
=\Bigl(\varphi_A
\Bigl[\begin{matrix}(\alpha)_3\\(\beta)_3\end{matrix}\Bigr]\Bigr)^2
\varphi_E
\Bigl[\begin{matrix}(\alpha)_3\\(\beta)_3\end{matrix}\Bigr]
\varphi_{\mathcal A}
\Bigl[\begin{matrix}(\alpha)_{12}\\(\beta)_{12}\end{matrix}\Bigr],\quad
\varphi^{\rm B}
\Bigl[\begin{matrix}\alpha\\\beta\end{matrix}\Bigr]
=\prod_i
\varphi^{\rm B}_{(\phi_i)}
\Bigl[\begin{matrix}\alpha\\\beta\end{matrix}\Bigr],
\end{align}
respectively, and also include the prefactor defined in subsection \ref{prefactor},
\begin{align}
\Phi
\Bigl[\begin{matrix}\alpha\\\beta\end{matrix}\Bigr]
=\varphi\Bigl[\begin{matrix}(\alpha)_{12}\\(\beta)_{12}\end{matrix}\Bigr]
\varphi^{\rm L}
\Bigl[\begin{matrix}(\alpha)_{12}\\(\beta)_{12}\end{matrix}\Bigr]
\varphi^{\rm B}\Bigl[\begin{matrix}\alpha\\\beta\end{matrix}\Bigr].
\end{align}
Note that the numerical factors 
$\varphi_A$, $\varphi_E$, $\varphi_{\mathcal A}$ and 
$\varphi^{\rm B}_{(\phi_i)}$ are only defined
for some special $(\alpha,\beta)$ sectors.
If we have not defined them in that sector, we simply regard them as
$1$.

The resulting factors $\Phi$ form an unwieldy matrix containing complex numbers.
In the $\beta=0$ sectors, these contributions are simply positive
integers and hence are easily interpreted as the number of fixed
points.
On the other hand, in the $\beta\ne 0$ sectors we encounter various
complex factors.
It is then natural to interpret them as fixed points with phase
contributions.
Namely, supposing we have $n_\alpha$ fixed points in the $\alpha$ sectors,
the numerical factor in each $(\alpha,\beta)$ sector should be
interpreted as
\begin{align}
\Phi
\Bigl[\begin{matrix}\alpha\\\beta\end{matrix}\Bigr]
=(e^{2\pi i\beta\varphi_1^{(\alpha)}}+e^{2\pi i\beta\varphi_2^{(\alpha)}}
+\cdots+e^{2\pi i\beta\varphi_{n_\alpha}^{(\alpha)}})e^{2\pi i\beta\varphi_0^{(\alpha)}},
\label{FPphase}
\end{align}
where each of the $n_\alpha$ fixed points acquires a phase
$\varphi_i^{(\alpha)}\;(i=1,2,\cdots,n_\alpha)$ in the orbifold action.
Here we have introduced the overall vacuum phase $\varphi_0^{(\alpha)}$ to make
the phases of the fixed points as simple as possible.
After separating the overall phase $\varphi_0^{(\alpha)}$, we find
that the remaining factors can be understood as the contributions 
from the fixed points with the definite phase $\varphi_i^{(\alpha)}$ 
as given in Table \ref{overallphase}.
\begin{table}[tb]
\begin{center}
\begin{tabular}{c||c|c|c|c|c|c}
$\alpha$&$1$&$2$&$3$&$4$&$5$&$6$ \\ \hline\hline
number of fixed points $n_\alpha$ &$1$&$1$&$1$&$3$&$1$&$2$ \\
phases of fixed points $\varphi^{(\alpha)}_i$ &$0$&$0$&$0$&$0,\pm1/4$&$0$&$\pm1/6$\\
overall phase $\varphi^{(\alpha)}_0$ &$19/48$&$-7/24$&$-1/16$&$1/12$&$23/48$&$-3/8$
\end{tabular}
\end{center}
\caption{Number of fixed points $n_\alpha$, 
the phases of fixed points $\varphi^{(\alpha)}_i$
and the overall phase $\varphi^{(\alpha)}_0$ in each sector. 
}
\label{overallphase}
\end{table}

\subsection{Phases from the lattice}\label{latphase}
Now let us turn to the physical part of the partition function.
Again, we focus only on the lattice part here, with the remainder considered in appendix \ref{Fermion}.
First, we consider the untwisted sector.
The original lattice before orbifolding is the
$[(A_2)^2\times(E_6)^3]\times[E_6]^*$ even self-dual lattice
with the conjugacy classes given in \eqref{OriginalLattice}.
For the $E_6$ part, the phases for the eigenstates of 
the permutation \eqref{EigenAdjointStates} are
given by $\omega^k$.
In addition, the shift\footnote{We have added subscript $E$ to
distinguish from the generator of the conjugacy classes of $A_2$, which
will also appear later.}
$(\bm\omega_E,\bm\omega_E,\bm\omega_E)/3$ introduces additional
phases.
The original $72$ states in each $E_6$ root lattice are
separated into $40,16$ and $16$ states with phases
$1,\omega$ and $\omega^2$, respectively, depending on their inner product with the shift.
Although at first sight all the $E_6$ gauge symmetries appear to be broken by the
shift, the diagonal symmetry is actually restored by combining the phases from
the permutation and shift.
Finally, the three sets of the 72 states acquire the phases
$1,\omega$ and $\omega^2$.

Next, we proceed to the twisted sectors.
Massless states and their phases
can be read
off from the physical part of the lattice partition functions
\eqref{Zfourtriplets} and \eqref{Z12LatticePF}.
The massless condition for the left-movers is given by\footnote{
As in section \ref{Sec:w/GeneralShift} and appendix \ref{Boson}, 
$(x)_1$ is the fractional part of $x$: $(x)_1 = x$ mod $1$.}
\begin{align}
\frac{1}{2}{\bm p}^2
+\Bigl(\frac{1}{2}\sum_{i}(\alpha\phi_i)_{1}
\bigl(1-(\alpha\phi_i)_{1}\bigr)-1\Bigr)=0,
\label{latmassless}
\end{align}
where the first term is the 
contribution from the diagonal $E_6$ and the two $A_2$ lattices,
${\bm p}^2
={\bm p_{E}}^2
+\sum_{j=1}^2 {\bm p_{Aj}}^2$, while the second term
represents the zero-point energy,
which gives $-1/3$ for $\alpha\ne 0\mbox{ mod }3$.
In the following, we shall focus only on the states that satisfy 
the above massless condition.

Let us start with the states of the $E_6$ part.
\begin{itemize}
\item
For the case of $\alpha\ne 0\mbox{ mod }3$, where
the partition function is given by \eqref{Eseparate}, 
the lattice momenta take values in the set
\begin{align}
S_E(\alpha)=\Bigl\{{\bm p_E}=\frac{1}{\sqrt{3}}
(\vec n\cdot\vec{\bm\alpha}_E+\alpha\bm\omega_E)\Big|\vec n\in{\mathbb Z}^6\Bigr\},
\end{align}
where $\bm\alpha_E$ and $\bm\omega_E$ are the simple roots and 
one of the fundamental weights corresponding to the generator of the conjugacy 
classes of $E_6$, respectively.
The lightest states for $\alpha=1$ and $2$ form the  ${\bf 27}$ and
$\overline{{\bf 27}}$ representations in $E_6$, respectively,
and their contribution to the mass is ${\bm p_E}^2/2=2/9$.
Since the partition function $e(\tau/3)$ takes the form
$27q^{2/9}+{\mathcal O}(q^{5/9})$, the next lightest states 
already exceed the massless condition.
\item
For the case of $\alpha=0\mbox{ mod }3$, the lattice partition
function ${\mathcal E}_k$ \eqref{calE} takes the form 
$(1+{\mathcal O}(q^1),{\mathcal O}(q^{4/3}),{\mathcal O}(q^{4/3}))$,
where the next lightest states in ${\mathcal O}(q^1)$ match the
massless condition exactly.
However, in Models 2 and 3, the $(A_2)^2$ part also gives a nonvanishing contribution,
and these states become massive.
Hence, only the origin can form massless states. 
\end{itemize}
We can read off the phases for these states from the partition function
in the $(\alpha,1)$ sector.
There are two types of phase contributions.
The first consists of the powers $\omega^{[\alpha/3]}$ in \eqref{physE}, while
the second originates from the partition functions
$e\bigl((\tau+1)/3\bigr)$ and $\omega^2e\bigl((\tau+2)/3\bigr)$ in
\eqref{physEmod3}.
Using the expansion of these partition functions ($\alpha=1,2$)
\begin{align}
e\Bigl(\frac{\tau+\alpha^{-1}}{3}\Bigr)
=\sum_{\bm p_E\in S_E(\alpha)}
e^{2\pi i\alpha^{-1}\bm p_E^2/2}q^{\bm p_E^2/2},
\end{align}
with $1^{-1}=1$ and $2^{-1}=2$ (mod 3) as before, we find that the second
contribution is given by $0,2/9,1/9$ for $\alpha=0,1,2\mbox{ mod }3$, respectively.
These phases for the lightest states are summarized in Table \ref{E6phase}.

\begin{table}
\begin{center}
\begin{tabular}{c||c|c|c|c|c|c|c}
$\alpha$&$0$&$1$&$2$&$3$&$4$&$5$&$6$\\
\hline\hline
phase&$0,0,1/3,2/3$&$2/9$&$1/9$&$1/3$&$5/9$&$4/9$&$2/3$
\end{tabular}
\end{center}
\caption{Phases originating from the $(E_6)^3$ lattice.
The first phase in the untwisted sector is the contribution from the
origin, while the remaining three are those from the three sets of 72
states after recombining the original 72 states in each $E_6$.}
\label{E6phase}
\end{table}

Next we consider the $(A_2)^2$ part, which also depends on the shift vectors
$(\bm\lambda^{}_{1},\bm\lambda^{}_{2})$.
\begin{itemize}
\item
For the sectors $\alpha\ne 0\mbox{ mod }3$,
whose partition function is given in \eqref{PhysA}, the lattice momenta can be read off as
\begin{align}
&\Bigl\{
\bm p_{A1}
=\vec n_1\cdot\vec{\bm\alpha}_A+\alpha\bm\lambda^{}_1+k_1\bm\omega_A
\Big|
\vec n_1\in{\mathbb Z}^2,k_1\in \{0,1,2\}
\Bigr\}\nonumber\\
&\otimes\Bigl\{
\bm p_{A2}
=\vec n_2\cdot\vec{\bm\alpha}_A+\alpha\bm\lambda^{}_2+k_2\bm\omega_A
\Big|\vec n_2\in{\mathbb Z}^2,k_2\in \{0,1,2\}
\Bigr\}, 
\end{align}
where $\bm\alpha_A$ and $\bm\omega_A$ are the simple roots and 
one of the fundamental weights corresponding to
the generator of the conjugacy classes of $A_2$, respectively.
After taking the mass contribution from the $E_6$ part into account, 
the massless states have to satisfy $\sum_{j=1}^2{\bm p_{Aj}}^2/2=1/9$.
\item
For the $\alpha=0\mbox{ mod }3$ sectors, since the origin is extracted
for both the $E_6$ part and the right-moving $[\widetilde A_2]^*$
part, we find that only the conjugacy class $(0,0,0,0)$
survives out of the whole set of conjugacy classes $\Pi$ \eqref{invset}.
Hence, only the states contributing to $A^{\vec l}_{k=0}$ 
in \eqref{Zfourtriplets} are relevant, whose momenta are given by
\begin{align}
\Bigl\{
\bm p_{A1}=\vec n_1\cdot\vec{\bm\alpha}_A
+\alpha\bm\lambda^{}_{1}\Big|\vec n_1\in{\mathbb Z}^2\Bigr\}
\otimes
\Bigl\{
\bm p_{A2}=\vec n_2\cdot\vec{\bm\alpha}_A
+\alpha\bm\lambda^{}_{2}\Big|\vec n_2\in{\mathbb Z}^2\Bigr\}.
\end{align}
As in the previous case, after substituting the mass contribution of the
$E_6$ part, the massless states have to satisfy
$\sum_{j=1}^2 {\bm p_{Aj}}^2/2=1$.
\end{itemize}
Again, the $(\alpha,1)$ sectors of the partition functions, \eqref{PhysA},
\eqref{PhysAdef} and \eqref{generalized}, imply 
the phase contribution
\begin{align}
\sum_{j=1}^2\Bigl(\frac{1}{2}\alpha^{-1}fk_j^2
-\frac{1}{2}\alpha(\bm\lambda^{}_{j})^2
+\bm p_{Aj}\circ\bm\lambda^{}_{j}\Bigr)
\end{align}
for each model (depending on $\bm\lambda_1$ and $\bm\lambda_2$).
Here $\alpha^{-1}$ denotes an integer satisfying $\alpha^{-1}\alpha=1$ mod 3 as
before, while in the case of $\alpha=0\mbox{ mod }3$ the first term
does not contribute.
We list these massless candidates in the $A_2$ part and their phase contributions for Models
2 and 3 in Tables \ref{Model2} and \ref{Model3}, respectively.

\begin{table}[htb]
\begin{center}
\begin{tabular}{c||c||c}
$\alpha$&state&phase\\
\hline\hline
0&
\begin{tabular}{c}
$|\pm\bm\alpha_1\rangle\otimes|\bm 0\rangle$\\
$(|\pm\bm\alpha_2\rangle,|\mp(\bm\alpha_1+\bm\alpha_2)\rangle)
\otimes|\bm 0\rangle$\\
$|\bm 0\rangle\otimes|\pm\bm\alpha_1\rangle$\\
$|\bm 0\rangle\otimes
(|\pm\bm\alpha_2\rangle,|\pm(\bm\alpha_1+\bm\alpha_2)\rangle)$
\end{tabular}
&
\begin{tabular}{c}
$\pm 1/6$\\
$\mp 1/12$\\
$0$\\
$\mp 1/4$
\end{tabular}
\\
\hline
1&\begin{tabular}{c}$-$\end{tabular}
&\begin{tabular}{c}$-$\end{tabular}
\\
\hline
2&
\begin{tabular}{c}
$|2\bm\lambda^{}_{1}\rangle
\otimes|2\bm\lambda^{}_{2}+\bm\omega-\bm\alpha_1-\bm\alpha_2\rangle$\\
$|2\bm\lambda^{}_{1}\rangle
\otimes|2\bm\lambda^{}_{2}-\bm\omega-\bm\alpha_2\rangle$  
\end{tabular}
&
\begin{tabular}{c}
$-4/9$\\
$1/18$ 
\end{tabular}
\\
\hline
3&
\begin{tabular}{c}
$(|3\bm\lambda^{}_{1}-\bm\alpha_1-\bm\alpha_2\rangle,
|3\bm\lambda^{}_{1}+\bm\alpha_2\rangle)
\otimes|3\bm\lambda^{}_{2}-\bm\alpha_1-2\bm\alpha_2\rangle$\\
$|3\bm\lambda^{}_{1}-\bm\alpha_1\rangle
\otimes(|3\bm\lambda^{}_{2}-\bm\alpha_1-\bm\alpha_2\rangle,
|3\bm\lambda^{}_{2}-\bm\alpha_2\rangle)$
\end{tabular}
&
\begin{tabular}{c}
$0$\\
$-1/3$
\end{tabular}
\\
\hline
4&
\begin{tabular}{c}
$(|4\bm\lambda^{}_{1}+\bm\omega-\bm\alpha_1\rangle,
|4\bm\lambda^{}_{1}-\bm\omega\rangle)
\otimes|4\bm\lambda^{}_{2}-\bm\alpha_1-2\bm\alpha_2\rangle$\\
$|4\bm\lambda^{}_{1}\rangle
\otimes|4\bm\lambda^{}_{2}-\bm\alpha_1-2\bm\alpha_2\rangle$  
\end{tabular}
&
\begin{tabular}{c}
$-17/36$\\
$5/18$
\end{tabular}
\\
\hline
5&
\begin{tabular}{c}
$(|5\bm\lambda^{}_{1}+\bm\omega-\bm\alpha_1\rangle,
|5\bm\lambda^{}_{1}-\bm\omega\rangle)
\otimes|5\bm\lambda^{}_{2}+\bm\omega-2\bm\alpha_1-3\bm\alpha_2\rangle$
\end{tabular}
&
\begin{tabular}{c}
$2/9$
\end{tabular}
\\
\hline
6&
\begin{tabular}{c}
$(|6\bm\lambda^{}_{1}-\bm\alpha_1-\bm\alpha_2\rangle,
|6\bm\lambda^{}_{1}+\bm\alpha_2\rangle)
\otimes(|6\bm\lambda^{}_{2}-2\bm\alpha_1-3\bm\alpha_2\rangle,
|6\bm\lambda^{}_{2}-\bm\alpha_1-3\bm\alpha_2\rangle)$\\
$|6\bm\lambda^{}_{1}-\bm\alpha_1\rangle
\otimes|6\bm\lambda^{}_{2}-2\bm\alpha_1-4\bm\alpha_2\rangle$\\
$|6\bm\lambda^{}_{1}-\bm\alpha_1\rangle
\otimes|6\bm\lambda^{}_{2}-\bm\alpha_1-2\bm\alpha_2\rangle$\\
$|6\bm\lambda^{}_{1}\rangle
\otimes|6\bm\lambda^{}_{2}-2\bm\alpha_1-4\bm\alpha_2\rangle$\\
$|6\bm\lambda^{}_{1}\rangle
\otimes|6\bm\lambda^{}_{2}-\bm\alpha_1-2\bm\alpha_2\rangle$ 
\end{tabular}
&
\begin{tabular}{c}
$-1/6$\\
$0$\\
$1/2$\\
$1/6$\\
$-1/3$
\end{tabular}
\end{tabular}
\end{center}
\caption{Massless candidates for Model 2 with
the shifts $\bm\lambda_1=\bm\alpha_1/12$, 
$\bm\lambda_2=(\bm\alpha_1+2\bm\alpha_2)/4$.
We have omitted the index $A$ in the simple roots $\bm\alpha_1$,
$\bm\alpha_2$ and the fundamental weight
$\bm\omega$.}
\label{Model2}
\end{table}

\begin{table}[htb]
\begin{center}
\begin{tabular}{c||c||c}
$\alpha$&state&phase\\
\hline\hline
0&
\begin{tabular}{c}
$|\pm\bm\alpha_1\rangle\otimes|\bm 0\rangle$\\
$|\pm\bm\alpha_2\rangle\otimes|\bm 0\rangle$\\
$|\mp(\bm\alpha_1+\bm\alpha_2)\rangle\otimes|\bm 0\rangle$\\
$|\bm 0\rangle\otimes|\pm\bm\alpha_1\rangle$\\
$|\bm 0\rangle\otimes
(|\pm\bm\alpha_2\rangle,|\mp(\bm\alpha_1+\bm\alpha_2)\rangle)$
\end{tabular}
&
\begin{tabular}{c}
$\mp 1/3$\\
$\mp 1/12$\\
$\pm 5/12$\\
$1/2$\\
$\mp 1/4$
\end{tabular}
\\
\hline
1&
\begin{tabular}{c}
$|\bm\lambda^{}_{1}+\bm\omega-\bm\alpha_1-\bm\alpha_2\rangle
\otimes|\bm\lambda^{}_{2}\rangle$
\end{tabular}
&
\begin{tabular}{c}
$1/9$ 
\end{tabular}
\\
\hline
2&
\begin{tabular}{c}
$|2\bm\lambda^{}_{1}-\bm\alpha_2\rangle
\otimes(|2\bm\lambda^{}_{2}+\bm\omega-\bm\alpha_1\rangle,
|2\bm\lambda^{}_{2}-\bm\omega\rangle)$ 
\end{tabular}
&
\begin{tabular}{c}
$1/18$ 
\end{tabular}
\\
\hline
3&
\begin{tabular}{c}
$|3\bm\lambda_1-2\bm\alpha_2\rangle\otimes|3\bm\lambda_2\rangle$\\
$|3\bm\lambda^{}_{1}-\bm\alpha_1-2\bm\alpha_2\rangle
\otimes|3\bm\lambda^{}_{2}\rangle$\\
$|3\bm\lambda^{}_{1}-\bm\alpha_2\rangle
\otimes(|3\bm\lambda^{}_{2}+\bm\alpha_2\rangle,
|3\bm\lambda_2-\bm\alpha_1-\bm\alpha_2\rangle)$
\end{tabular}
&
\begin{tabular}{c}
$0$\\
$1/3$\\
$-1/3$
\end{tabular}
\\
\hline
4&
\begin{tabular}{c}
$|4\bm\lambda^{}_{1}+\bm\omega-\bm\alpha_1-2\bm\alpha_2\rangle
\otimes|4\bm\lambda^{}_{2}-\bm\alpha_1\rangle$\\
$|4\bm\lambda^{}_{1}-\bm\omega-2\bm\alpha_2\rangle
\otimes|4\bm\lambda^{}_{2}-\bm\alpha_1\rangle$\\
$|4\bm\lambda^{}_{1}-2\bm\alpha_2\rangle
\otimes|4\bm\lambda^{}_{2}-\bm\alpha_1\rangle$
\end{tabular}
&
\begin{tabular}{c}
$-17/36$\\
$1/36$\\
$-2/9$
\end{tabular}
\\
\hline
5&
\begin{tabular}{c}
$|5\bm\lambda^{}_{1}+\bm\omega-\bm\alpha_1-3\bm\alpha_2\rangle
\otimes|5\bm\lambda^{}_{2}-\bm\alpha_1\rangle$
\end{tabular}
&
\begin{tabular}{c}
$2/9$
\end{tabular}
\\
\hline
6&
\begin{tabular}{c}
$|6\bm\lambda^{}_{1}-\bm\alpha_1-4\bm\alpha_2\rangle
\otimes|6\bm\lambda^{}_{2}-2\bm\alpha_1\rangle$,
$|6\bm\lambda^{}_{1}-2\bm\alpha_2\rangle
\otimes|6\bm\lambda^{}_{2}-\bm\alpha_1\rangle$\\
$|6\bm\lambda^{}_{1}-\bm\alpha_1-4\bm\alpha_2\rangle
\otimes|6\bm\lambda^{}_{2}-\bm\alpha_1\rangle$,
$|6\bm\lambda^{}_{1}-2\bm\alpha_2\rangle
\otimes|6\bm\lambda^{}_{2}-2\bm\alpha_1\rangle$\\
$|6\bm\lambda^{}_{1}-\bm\alpha_1-3\bm\alpha_2\rangle
\otimes(|6\bm\lambda^{}_{2}-2\bm\alpha_1-\bm\alpha_2\rangle,
|6\bm\lambda^{}_{2}-\bm\alpha_1+\bm\alpha_2\rangle)$\\
$|6\bm\lambda^{}_{1}-3\bm\alpha_2\rangle
\otimes(|6\bm\lambda^{}_{2}-2\bm\alpha_1-\bm\alpha_2\rangle,
|6\bm\lambda^{}_{2}-\bm\alpha_1+\bm\alpha_2\rangle)$
\end{tabular}
&
\begin{tabular}{c}
$1/3$\\
$-1/6$\\
$1/2$\\
$1/6$
\end{tabular}
\end{tabular}
\end{center}
\caption{Massless candidates for Model 3 with
the shifts $\bm\lambda_1=(\bm\alpha_1+6\bm\alpha_2)/12$,
$\bm\lambda_2=\bm\alpha_1/4$.}
\label{Model3}
\end{table}

\subsection{Massless spectrum}
In the previous subsections, we calculated the phases that are relevant 
to the massless spectra of Models 2 and 3.
It is now necessary to combine them to form phaseless
states.
Let us examine how an adjoint Higgs field and chiral (antichiral)
generations appear in Model 2 as an example.
Note that, in each of the $\alpha=0$ and $6$ sectors, massless states and
their CPT conjugate states exist in the same sector and compose the
untwisted sector $U$ and twisted sector $T_6$.
In the other sectors, states and their CPT conjugate states reside in the
$\alpha$ and $12-\alpha$ sectors, respectively, and compose the twisted
sectors $T_{1,2,3,4,5}$. 

We first consider the untwisted sector $U$.
This sector contains gauge fields and the adjoint Higgs field.
In Model 2, in addition to the diagonal $E_6$,
a non-Abelian part from one of the $(A_2)^2$ survives as seen in Table \ref{Model2}.
Adding the Abelian parts from excited bosons with vanishing phases, the gauge group
of Model 2 turns out to be
\begin{align}
{\rm Model}\ 2:(E_6)_3\times SU(2)\times U(1)^3,
\end{align}
where the lower index denotes the Kac-Moody level of the gauge group.
The other states originating from $(E_6)^3$ acquire phase contributions
$\omega^{\pm 1}$ as discussed at the beginning of subsection
\ref{latphase}, which are canceled by the phases originating from the
right-moving states $|0,0,0,\pm 1\rangle$ (NS) or
$|\pm(1/2,-1/2,-1/2,1/2)\rangle$ (R) to form massless fields 
(See Table \ref{fermionphase} in appendix \ref{Fermion}.).
We define the four-dimensional chirality as `left-handed' if the first
component of the fermionic states is $1/2$.
Since these fields are in the adjoint representation of $(E_6)_3$ gauge
symmetry and do not have nontrivial charges in the other gauge
symmetries, they compose the left-handed chiral multiplet
$({\bf 78},{\bf 1},0,0,0)_{{\rm L}}$ of $E_6\times SU(2)\times U(1)^3 $.
There are also chiral multiplets in the nontrivial representation of
$U(1)^3$, $({\bf 1},{\bf 1},+6,\pm 3,0)_{{\rm L}}$, where the three
$U(1)$ are normalized with the unit 
$(\sqrt{2}/12,\sqrt{6}/6,\sqrt{6}/12 )$. 

Let us proceed to the twisted sectors $T_{\alpha}$.
As we have seen in Table \ref{Model2}, there are no massless candidates
in the $\alpha =1$ sector and therefore no massless
fields in the twisted sector $T_1$.
In the $\alpha=2$ twisted sector in Table \ref{Model2}, there are two massless candidates,
$|2\bm\lambda^{}_{1}\rangle
\otimes|2\bm\lambda^{}_{2}+\bm\omega-\bm\alpha_1-\bm\alpha_2\rangle$
and
$|2\bm\lambda^{}_{1}\rangle\otimes
|2\bm\lambda^{}_{2}-\bm\omega-\bm\alpha_2\rangle$.
Combined with the lightest momentum states in $S_E(2)$, 
which correspond to the $\overline{{\bf 27}}$ representation,
and the right-moving fermionic states,
only the latter candidate cancels the phase and survives after the projection.
Then, with its CPT conjugate in the $\alpha=10$ sector,
it composes 
the multiplet
$(\overline{{\bf 27}},{\bf 1},+2,0,-2 )_{{\rm L}}$.
A similar analysis can be performed for the other sectors.
Note that in the twisted sectors $T_4$ and $T_6$, we have to take into
account the fixed points (three fixed points for $T_4$ and two fixed
points for $T_6$) and their phases in Table \ref{overallphase}.
In this way, one can find all the massless fields in
Model 2 and also those in other models.

We list the resulting spectra of the three models with three
generations in Table \ref{spectra}.
We omit the gauge and gravity multiplets in the table. 
The gauge group of Models 1 and 2 is
$E_6\times SU(2)\times U(1)^3$ and that of Model 3 is
$E_6\times U(1)^4$.
Each model contains a chiral multiplet in the adjoint representation
of the level $3$ $E_6$ group which corresponds to a GUT adjoint Higgs
field.
It turns out that the numbers of chiral and antichiral generations
for Models 1 and 3 are $5$ and $2$, while they are $4$ and $1$ for Model
2, respectively.
Hence, each model leads to a net of three chiral generations.
Models 1 and 2 contain a hidden gauge group $SU(2)$ and its
doublet field in the twisted sector $T_6$, while there is no
non-Abelian hidden sector in Model 3.
As mentioned in the previous section, the massless spectrum of Model 1 is the same as that
analyzed in the framework of the ${\mathbb Z}_6$ orbifold model
\cite{KTE6}.
Although it is possible that the two models,
which are constructed in ${\mathbb Z}_6$ and ${\mathbb Z}_{12}$, respectively, 
have different interactions, they are likely to be the same.
On the other hand, the other two models, Models 2 and 3, are completely new.

\begin{table}[htb] 
\begin{center}
\begin{tabular}{ccccc}
\hline\hline
& Model 1 & Model 2 & Model 3 \\ 
\hline
\begin{tabular}{c}gauge\\symmetry\end{tabular}
& $E_6 \times SU(2) \times U(1)^3 $ 
& $E_6 \times SU(2) \times U(1)^3 $
& $E_6 \times U(1)^4$ \\
\hline
$U$ & \begin{tabular}{c}
$({\bf 78},{\bf 1},0,0,0)_{{\rm L}}$\\
$({\bf 1},{\bf 1},+6,0,0)_{{\rm L}}$
      \end{tabular}
    & \begin{tabular}{c}
$({\bf 78},{\bf 1},0,0,0)_{{\rm L}}$\\
$({\bf 1},{\bf 1},+6,\pm 3,0)_{{\rm L}}$
      \end{tabular}
    & \begin{tabular}{c}
$({\bf 78},0,0,0,0)_{{\rm L}}$\\
$({\bf 1},-6,0,0,0)_{{\rm L}}$\\
$({\bf 1},+3,\pm 6,0,0)_{{\rm L}}$
      \end{tabular}
\\ \phantom{,}\vspace{-4.3mm} \\
$T_1$ & $({\bf 27},{\bf 1},+1,0,\pm 1)_{{\rm L}}$ 
      & --- 
      & \phantom{,} $({\bf 27},-1,-1,+1,0)_{{\rm L}}$ \phantom{,}  
\\ \phantom{,}\vspace{-4.3mm} \\
$T_2$ & $(\overline{{\bf 27}},{\bf 1},-1,\pm 1,0)_{{\rm L}}$ 
      & $(\overline{{\bf 27}},{\bf 1},+2,0,-2)_{{\rm L}}$ 
      & $(\overline{{\bf 27}},+1,0,0,\pm 1)_{{\rm L}}$   
\\ \phantom{,}\vspace{-4.3mm} \\
$T_3$ & \begin{tabular}{c} $2({\bf 1},{\bf 1},-3,0,\pm 3)_{{\rm L}}$ 
        \end{tabular}
      & \begin{tabular}{c} $({\bf 1},{\bf 1},-3,\pm 3,-3)_{{\rm L}}$ 
        \end{tabular}
      & \begin{tabular}{c} $({\bf 1},+3,-3,+3,0)_{{\rm L}}$  \\ 
                           $({\bf 1},+3,+3,-3,0)_{{\rm L}}$
        \end{tabular}
\\ \phantom{,}\vspace{-4.3mm} \\
$T_4$ & \begin{tabular}{c} $({\bf 27},{\bf 1},-2,0,0)_{{\rm L}}$
        \end{tabular}
      & \begin{tabular}{c} $({\bf 27},{\bf 1},-2,\pm 1,0)_{{\rm L}}$
        \end{tabular}
      & \begin{tabular}{c} $({\bf 27},+2,0,0,0)_{{\rm L}}$ \\
                           $({\bf 27},-1,\pm 2,0,0)_{{\rm L}}$
        \end{tabular}
\\ \phantom{,}\vspace{-4.3mm} \\
$T_5$ & $({\bf 27},{\bf 1},+1,0,\pm 1)_{{\rm L}}$ 
      & $({\bf 27},{\bf 1},+1,\pm 1,+1)_{{\rm L}}$ 
      &  $({\bf 27},-1,+1,-1,0)_{{\rm L}}$  
\\ \phantom{,}\vspace{-4.3mm} \\
$T_6$ & \begin{tabular}{c} $({\bf 1},{\bf 2},0,0,\pm 3)_{{\rm L}}$ \\
                           $({\bf 1},{\bf 1},+3,\pm 3,0)_{{\rm L}}$
        \end{tabular}
      & \begin{tabular}{c} $({\bf 1},{\bf 2},0,\pm 3,0)_{{\rm L}}$ \\
                           $({\bf 1},{\bf 1},-6,0,+6)_{{\rm L}}$
        \end{tabular}
      & \begin{tabular}{c} $({\bf 1},-3,0,0,\pm 3)_{{\rm L}}$ \\
                           $({\bf 1},0,+6,-2,0)_{{\rm L}}$ \\
                           $({\bf 1},0,-6,+2,0)_{{\rm L}}$
        \end{tabular} \\
\hline
\begin{tabular}{c}normalization\\of $U(1)$\end{tabular}
&$\Bigl(\frac{\sqrt{2}}{6},\frac{\sqrt{6}}{6},\frac{\sqrt{6}}{6}\Bigr)$
&$\Bigl(\frac{\sqrt{2}}{12},\frac{\sqrt{6}}{6},\frac{\sqrt{6}}{12}\Bigr)$
&$\Bigl(\frac{\sqrt{2}}{6},\frac{\sqrt{6}}{12},
\frac{\sqrt{2}}{4},\frac{\sqrt{6}}{6}\Bigr)$\\
\hline\hline
\end{tabular}
\end{center}
\caption{Massless spectra of the models with three generations:
$U$ and $T_\alpha$ denote the untwisted and various twisted sectors,
respectively.
The quantum numbers of the left-handed chiral multiplets and the
normalizations of the $U(1)$ charges are shown.
The irrational normalizations of the $U(1)$ originate from a general
decomposition of the Lie algebra into its subalgebras.
The gravity and gauge multiplets are omitted.}
\label{spectra}
\end{table}

\section{Summary and discussion}
As we reported briefly in \cite{letter}, we have found two novel
four-dimensional ${\mathcal N}=1$ $E_6$ grand unified models with an
adjoint Higgs field with three generations in the framework of the
asymmetric orbifold of heterotic string theory.
Before this work, only one such $E_6$ unified model was known,
which was claimed to be unique in the classification \cite{KTE6}.

In this paper, we have presented all the details and techniques
used in our construction, in the hope that
they will be useful in the construction of other models using
heterotic string theory.
We would like to stress that, with all the techniques collected from
previous works, one can now systematically design the setup of
heterotic string theory to satisfy various requirements at will.

Actually, one of our motivations in this work was to embed 
the scenario of the anomalous $U(1)$ GUT
into the framework of string theory.
Unfortunately, similarly to the model in \cite{KTE6}, we found that our new
models do not possess additional gauge symmetries, such
as the anomalous $U(1)_A$ gauge symmetry
\cite{anomalousU(1),anomalousU(1)ph,anomalousU(1)GUT} and
$SU(2)_F$ family symmetry \cite{horizontal,SCPV}, which help to prevent
the doublet-triplet splitting problem and the SUSY flavor/CP problem.
Our models also share the property that a $({\mathbb Z}_3)^3$
subgroup of the $U(1)^3$ symmetry remains unbroken even after all the
singlets develop nonvanishing vacuum expectation values.
Nevertheless, our discovery of new models that have been missed 
from the classification raises hopes for the discovery of
many other new models including phenomenologically desirable ones.

In the rest of this section, we discuss some related issues 
on our formulation of the
partition function and the interpretation of the orbifold projection.

In \cite{KTE6} a similar argument using the modular invariant partition
function was presented with the concept of the conjugacy classes defined
by modding out the dual invariant sublattice by the invariant
sublattice.
It is in general, however, difficult to find the conjugacy classes
explicitly in this formulation, particularly when orbifolds with
permutation are considered.
Our formulation is based on the conjugacy classes of the Lie algebra.
Hence, we can always write down the formula explicitly without
difficulty.
Also note that our formulation is applicable to any of the ${\mathbb Z}_n$
actions, although the explicit expression depends on the details of the
orbifold actions.

Finally, we comment on the assignment of phases to the massive 
part of the $T_6$ twisted sector of the lattice partition function.
In the $(6,\beta)$ sectors, we have two different lattices depending on the 
value of $\beta$, as shown in \eqref{tildeA} and \eqref{tildeAdef}.
The $\beta=0,6$ sectors with an argument of $\tau/2$ correspond
to a condensed $A_2$ lattice with root length $\sqrt{2}\times
1/\sqrt{2}=1$, while the $\beta=3,9$ sectors with an argument of $2\tau$
correspond to a dilute $A_2$ lattice with root length
$\sqrt{2}\times\sqrt{2}=2$.
Although it is possible to assign phases for these sectors so that 
the contributions from the extra lattice points in the condensed lattice 
cancel among themselves to give a dilute lattice in the $\beta=3,9$ 
sectors, we cannot identify a unique phase assignment because its 
interpretation in terms of shifts or twists is not clear.
Nevertheless, our massless spectra do not depend on how the phase assignment is chosen, 
since only the origin of the right-moving lattice contributes 
to the massless states.
It, however, will be interesting to study how the assignment is fixed.
We hope to return to this point in our future work.

\section*{Acknowledgments}
We appreciate T.~Arai, K.~Hosomichi, H.~Kanno, Y.~Kazama, Y.~Kawamura,
T.~Kobayashi, J.~C.~Lee, S.~Mizoguchi, H.~Nakano, Y.~Sugawara and
T.~Takahashi for valuable discussions.
This work was partially supported by Grants-in-Aid for the Nagoya
University Global COE Program (G07) and Scientific Research
on Priority Areas [\#22011004] (N.M.) and for Young
Scientists (B) [\#21740176] (S.M.) from the Ministry of Education,
Culture, Sports, Science and Technology of Japan.
The work of T.Y. was partially supported by the Japanese Society for
the Promotion of Science.

\appendix

\section{Partition functions of fermionic/bosonic oscillators}\label{osc_pf}
The Dedekind eta function and Jacobi theta function are respectively defined as
\begin{align}
\eta(\tau)&=q^{1/24}\prod_{m=1}^\infty(1-q^m),\nonumber\\
\vartheta\Bigl[\begin{matrix}\alpha\\\beta\end{matrix}\Bigr]
&=\eta(\tau)q^{\alpha^2/2-1/24}e^{2\pi i\alpha\beta}
\prod_{m=1}^\infty
(1+q^{m+\alpha-1/2}e^{2\pi i\beta})
(1+q^{m-\alpha-1/2}e^{-2\pi i\beta})
\nonumber\\
&=\sum_ne^{-\pi(n+\alpha)(-i\tau)(n+\alpha)+2\pi i(n+\alpha)\beta}.
\end{align} 
Using the Poisson resummation formula, the modular transformations of 
these partition functions are as follows:
\begin{align}
\eta(\tau+1)&=e^{2\pi i(1/24)}\eta(\tau),
&\eta(-1/\tau)&=\sqrt{-i\tau}\eta(\tau),\nonumber\\
\vartheta\Bigl[\begin{matrix}\alpha\\\beta\end{matrix}\Bigr](\tau+1)
&=e^{\pi i\alpha(1-\alpha)}\vartheta
\Bigl[\begin{matrix}\alpha\\\beta+\alpha-1/2\end{matrix}\Bigr](\tau),
&\vartheta\Bigl[\begin{matrix}\alpha\\\beta\end{matrix}\Bigr](-1/\tau)
&=\sqrt{-i\tau}e^{2\pi i\alpha\beta}\vartheta
\Bigl[\begin{matrix}\beta\\-\alpha\end{matrix}\Bigr](\tau).
\end{align}

\subsection{Fermion}
Let us define the fermion partition function by
\begin{align}
\Theta^{\rm F}_{(\phi)}
\Bigl[\begin{matrix}\alpha\\\beta\end{matrix}\Bigr]
=\frac{e^{-\pi i(\alpha\phi)\cdot(\beta\phi)}}{2\eta^4}\biggl\{
\prod_{i=0}^3\vartheta
\Bigl[\begin{matrix}\alpha\phi_i\\\beta\phi_i\end{matrix}\Bigr]
-\prod_{i=0}^3e^{-\pi i\alpha\phi_i}\vartheta
\Bigl[\begin{matrix}\alpha\phi_i\\\beta\phi_i+1/2\end{matrix}\Bigr]
-\prod_{i=0}^3\vartheta
\Bigl[\begin{matrix}\alpha\phi_i+1/2\\\beta\phi_i\end{matrix}\Bigr]
\biggr\},
\label{fermionPF}
\end{align}
with $2\pi\phi_i$ $(i=0,1,2,3)$ being the rotation angles of a
${\mathbb Z}_N$ orbifold action on right-moving transverse complex
four-dimensional space.
In addition to the orbifold condition
\begin{align}
\phi_0=0,\quad N\phi_i\in{\mathbb Z},
\end{align}
the $\phi_i$ have to satisfy the fermion consistency condition
\begin{align}
\sum_iN\phi_i=0\quad\mbox{mod 2}.
\label{orbifoldtwist}
\end{align}
Here the first two terms correspond to the GSO-projected NS sector,
while the third term corresponds to the GSO-projected R sector.

This partition function transforms under the modular transformation as
\begin{align}
\Theta^{\rm F}_{(\phi)}
\Bigl[\begin{matrix}\alpha\\\beta\end{matrix}\Bigr](\tau+1)
=e^{-2\pi i(16/24)}\Theta^{\rm F}_{(\phi)}
\Bigl[\begin{matrix}\alpha\\\beta+\alpha\end{matrix}\Bigr](\tau),\quad
\Theta^{\rm F}_{(\phi)}
\Bigl[\begin{matrix}\alpha\\\beta\end{matrix}\Bigr](-1/\tau)
=\Theta^{\rm F}_{(\phi)}
\Bigl[\begin{matrix}\beta\\-\alpha\end{matrix}\Bigr](\tau).
\end{align}
It also satisfies the periodicity relation
\begin{align}
\Theta^{\rm F}_{(\phi)}
\Bigl[\begin{matrix}\alpha\\\beta+N\end{matrix}\Bigr]
=e^{\pi i\alpha\phi\cdot N\phi}\Theta^{\rm F}_{(\phi)}
\Bigl[\begin{matrix}\alpha\\\beta\end{matrix}\Bigr]
\label{fermionperiod}
\end{align}
if we impose the fermion consistency condition
\eqref{orbifoldtwist}.
Since we are imposing the SUSY condition
\begin{align}
\sum_i\phi_i=0\quad\mbox{mod 2},
\end{align}
the above consistency condition \eqref{orbifoldtwist} is automatically satisfied.

\subsection{Boson}\label{Boson}
For every complex dimension, the twisted boson partition function is given by
\begin{align}
\Theta^{\rm B}_{(\phi)}
\Bigl[\begin{matrix}\alpha\\\beta\end{matrix}\Bigr]
=e^{i\pi\alpha\phi(\beta\phi-1)}\frac{\eta}
{\vartheta\Bigl[\begin{matrix}\alpha\phi+1/2\\
\beta\phi-1/2\end{matrix}\Bigr]},
\label{bosonPF}
\end{align}
while the untwisted boson partition function is  $1/\eta^2$.
Under the modular transformations, it satisfies
\begin{align}
\Theta^{\rm B}_{(\phi)}
\Bigl[\begin{matrix}\alpha\\\beta\end{matrix}\Bigr](\tau+1)
=e^{-2\pi i(2/24)}\Theta^{\rm B}_{(\phi)}
\Bigl[\begin{matrix}\alpha\\\beta+\alpha\end{matrix}\Bigr](\tau),\quad
\Theta^{\rm B}_{(\phi)}
\Bigl[\begin{matrix}\alpha\\\beta\end{matrix}\Bigr](-1/\tau)
=i\Theta^{\rm B}_{(\phi)}
\Bigl[\begin{matrix}\beta\\-\alpha\end{matrix}\Bigr](\tau).
\label{i}
\end{align}
Note that an extra phase $i$ appears in the 
${\mathcal S}$-transformation compared with the untwisted boson
partition function $1/\eta^2(\tau)$.
Therefore, to simplify the modular transformation of the total partition function,
we define the lattice partition function in the text so that it
transforms with the same factor $i$ for every complex dimension.

The orbifold periodic relation is given by
\begin{align}
\Theta^{\rm B}_{(\phi)}
\Bigl[\begin{matrix}\alpha\\\beta+N\end{matrix}\Bigr]
=e^{-\pi i\alpha\phi N\phi}e^{-\pi iN\phi}\Theta^{\rm B}_{(\phi)}
\Bigl[\begin{matrix}\alpha\\\beta\end{matrix}\Bigr].
\label{bosonperiod}
\end{align}
In the right-moving case, since we have the consistency condition
\eqref{orbifoldtwist}, the second factor $e^{-\pi iN\phi}$ is cancelled.
Combined with the result for the fermion partition function
\eqref{fermionperiod}, it implies that
the phase contributions of the right-moving modes to the total orbifold 
periodic condition (\ref{orbperiod}) always cancel among themselves.

Although the above boson partition function is defined to have a desirable
modular transformation property,
it does not take a form suitable for physical
interpretation.
For this reason we rewrite the partition
function $\Theta^{\rm B}_{(\phi)}$ as the product of the physical
partition function $\widehat\Theta^{\rm B}_{(\phi)}$ and an
overall multiplicative factor 
$\varphi^{\rm B}_{(\phi)}$:
\begin{align}
\Theta^{\rm B}_{(\phi)}
\Bigl[\begin{matrix}\alpha\\\beta\end{matrix}\Bigr]
&=\varphi^{\rm B}_{(\phi)}
\Bigl[\begin{matrix}\alpha\\\beta\end{matrix}\Bigr]
\widehat\Theta^{\rm B}_{(\phi)}
\Bigl[\begin{matrix}\alpha\\\beta\end{matrix}\Bigr].
\label{bosonseparate}
\end{align}
The factor $\varphi^{\rm B}_{(\phi)}$ and the physical partition
function $\widehat\Theta^{\rm B}_{(\phi)}$ are defined as
\begin{align}
\varphi^{\rm B}_{(\phi)}
\Bigl[\begin{matrix}\alpha\\\beta\end{matrix}\Bigr]
&=(-i)(-1)^{[\alpha\phi]}
e^{-\pi i\beta\phi(\alpha\phi-1)+2\pi i\beta\phi[\alpha\phi]},\nonumber\\
\widehat\Theta^{\rm B}_{(\phi)}
\Bigl[\begin{matrix}\alpha\\\beta\end{matrix}\Bigr]
&=\frac{1}{q^{(1/2-(\alpha\phi)_{1})^2/2-1/24}
\prod_{n=1}^\infty(1-q^{n-1+(\alpha\phi)_{1}}e^{2\pi i\beta\phi})
(1-q^{n-(\alpha\phi)_{1}}e^{-2\pi i\beta\phi})}
\label{noninteger}
\end{align}
for $\alpha\phi\not\in{\mathbb Z}$ and
\begin{align}
\varphi^{\rm B}_{(\phi)}
\Bigl[\begin{matrix}\alpha\\\beta\end{matrix}\Bigr]
&=\frac{(-1)^{\alpha\phi}e^{\pi i\beta\phi\alpha\phi}}
{2\sin\beta\pi\phi},\nonumber\\
\widehat\Theta^{\rm B}_{(\phi)}
\Bigl[\begin{matrix}\alpha\\\beta\end{matrix}\Bigr]
&=\frac{1}{q^{2/24}\prod_{n=1}^\infty
(1-q^ne^{2\pi i\beta\phi})(1-q^ne^{-2\pi i\beta\phi})}
\label{integer}
\end{align}
for $\alpha\phi\in{\mathbb Z}$, $\beta\phi\not\in{\mathbb Z}$.
Here $[x]$ is the largest integer that does not
exceed $x$, $[x]\le x<[x]+1$, while $(x)_{1}$ is defined as $x$ mod 1,
$(x)_{1}=x-[x]$.

\subsection{Phases from fermions}\label{Fermion}
Here we summarize the orbifold phases originating from the fermions 
by focusing on the massless states.
For this purpose, it is easiest to view the fermionic state as the
lattice state on the $D_4=SO(8)$ lattice.
Before orbifolding,
the states in the NS sector are given by the $D_4$ root lattice
shifted by the weight of the vector representation ${\bf 8}_v$, while
the states in the R sector are given by the lattice shifted by that of
the spinor representation ${\bf 8}_s$.
Namely, they are given by
\begin{align}
\mbox{NS:}\;\Bigl\{\Big|n_0+1,n_1,n_2,n_3\Big\rangle\Bigr\},\quad
\mbox{R:}\;\Bigl\{\Big|n_0+\frac{1}{2},
n_1+\frac{1}{2},n_2+\frac{1}{2},n_3+\frac{1}{2}\Big\rangle\Bigr\},
\end{align}
where $n_i\in{\mathbb Z}$ are subject to the constraint
$\sum_{i=0}^3n_i\in 2{\mathbb Z}$.
Under the twist action, these lattice states are further
shifted by $\alpha\phi_i$ in the $\alpha$ sector.
For the lattice point $|s_0,s_1,s_2,s_3\rangle$, the mass contribution
from the fermion partition function can be read off from
\eqref{fermionPF}:
$q^{-4/24+\sum_{i=0}^3(s_i)^2/2}=q^{\sum_{i=0}^3\{(s_i)^2/2-1/24\}}$,
although it has to be supplemented by the mass contribution from the
vacuum state of the boson partition function as given in
\eqref{noninteger} and \eqref{integer}:
$q^{-\sum_{i=0}^3\{(1/2-(\alpha\phi_i)_1)^2/2-1/24\}}$. 
Therefore, the massless condition is simply
\begin{align}
\frac{1}{2}\sum_{i=0}^3(s_i)^2
+\frac{1}{2}\sum_{i=0}^3(\alpha\phi_i)_1\bigl(1-(\alpha\phi_i)_1\bigr)
-\frac{1}{2}=0.
\label{massless}
\end{align}
Here we have considered only the ground state, 
since the contribution from the bosonic oscillators makes
the states massive.
The phases obtained by the orbifold action $\theta^{\beta=1}$ for the
$\alpha$-twisted massless states can be read off from the partition
function \eqref{fermionPF} as
\begin{align}
\frac{\alpha}{2}\sum_{i=0}^3(\phi_i)^2-\sum_{i=0}^3s_i\phi_i,
\end{align}
where we include the minus sign originating from the complex conjugate of
the right-moving part.
We list these massless states and their phase contributions in Table
\ref{fermionphase}.

\begin{table}[htb]
\begin{center}
\begin{tabular}{c||c|c||c}
$\alpha$&NS&R&phase\\
\hline\hline
0&
\begin{tabular}{c}
$|\pm 1,0,0,0\rangle$\\
$|0,\pm 1,0,0\rangle$\\
$|0,0,\pm 1,0\rangle$\\
$|0,0,0,\pm 1\rangle$
\end{tabular}
&
\begin{tabular}{c}
$|\pm(1/2,1/2,1/2,1/2)\rangle$\\
$|\pm(1/2,1/2,-1/2,-1/2)\rangle$\\
$|\pm(1/2,-1/2,1/2,-1/2)\rangle$\\
$|\pm(1/2,-1/2,-1/2,1/2)\rangle$
\end{tabular}
&
\begin{tabular}{c}
$0$\\
$\mp 1/12$\\
$\pm 5/12$\\
$\mp 1/3$
\end{tabular}
\\
\hline
1&$|0,1/12,7/12,1/3\rangle$&$|1/2,-5/12,1/12,-1/6\rangle$&$13/48$\\
\hline
2&$|0,1/6,1/6,2/3\rangle$&$|1/2,-1/3,-1/3,1/6\rangle$&$1/8$\\
\hline
3&
\begin{tabular}{c}
$|0,-3/4,-1/4,0\rangle$\\
$|0,1/4,3/4,0\rangle$
\end{tabular}
&
\begin{tabular}{c}
$|-1/2,-1/4,1/4,1/2\rangle$\\
$|1/2,-1/4,1/4,-1/2\rangle$
\end{tabular}
&
\begin{tabular}{c}
$19/48$\\
$-13/48$
\end{tabular}
\\
\hline
4&$|0,1/3,1/3,1/3\rangle$&$|1/2,-1/6,-1/6,-1/6\rangle$&$-5/12$\\
\hline
5&$|0,-7/12,-1/12,-1/3\rangle$&$|-1/2,-1/12,5/12,1/6\rangle$&$-7/48$\\
\hline
6&
\begin{tabular}{c}
$|0,-1/2,-1/2,0\rangle$\\
$|0,1/2,1/2,0\rangle$
\end{tabular}
&
\begin{tabular}{c}
$|-1/2,0,0,1/2\rangle$\\
$|1/2,0,0,-1/2\rangle$
\end{tabular}
&
\begin{tabular}{c}
$-7/24$\\
$1/24$
\end{tabular}
\end{tabular}
\end{center}
\caption{Massless states and the corresponding phases for the
fermions.}
\label{fermionphase}
\end{table}

\section{Lattice decomposition}\label{lattice_decomp}
\subsection{$E_8\to E_6\times A_2$}
\begin{figure}[htb]
 \begin{center}
\scalebox{0.4}[0.4]{\includegraphics{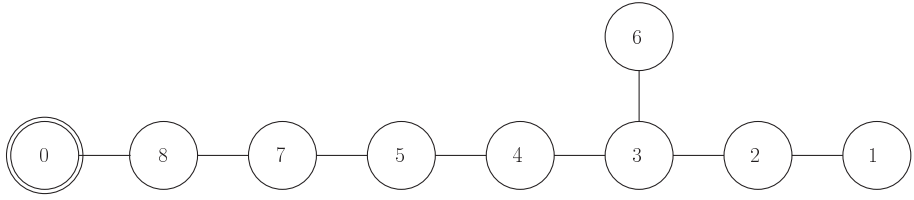}}
 \end{center}
 \caption{Dynkin diagram of the $E_8$ group.}
 \label{fig:E8Dynkin}
 \end{figure}

We begin with the decomposition of the $E_8$ lattice into the $E_6$ and $A_2$ lattices. 
As is well known, the $E_8$ lattice is the only even self-dual lattice 
in the Euclidean eight-dimensional space.
The $E_8$ root lattice is constructed from the sum of the multiples of
its simple roots ${\bm\alpha}_1,\cdots,{\bm\alpha}_8$.
The Dynkin diagram of the $E_8$ group is depicted in Figure \ref{fig:E8Dynkin}.
According to the general theory of Lie algebras, a maximal subalgebra can be 
obtained by adding one more node ${\bm\alpha}_0$ to form an extended Dynkin diagram
and subsequently removing one of its nodes.
In the case of $E_8$, the minimal root to be added is
\begin{align}
{\bm\alpha}_0=-2{\bm\alpha}_1-4{\bm\alpha}_2-6{\bm\alpha}_3
-5{\bm\alpha}_4-4{\bm\alpha}_5-3{\bm\alpha}_7-2{\bm\alpha}_8
-3{\bm\alpha}_6.
\end{align}
For the decomposition into $E_6\times A_2$, we have to remove
${\bm\alpha}_7$, and identify the remaining roots with those for $E_6\times A_2$ as follows:
\begin{align}
&E_6:\quad
{\bm\alpha}'_1={\bm\alpha}_1,\quad
{\bm\alpha}'_2={\bm\alpha}_2,\quad
{\bm\alpha}'_3={\bm\alpha}_3,\quad
{\bm\alpha}'_4={\bm\alpha}_4,\quad
{\bm\alpha}'_5={\bm\alpha}_5,\quad
{\bm\alpha}'_6={\bm\alpha}_6,\\
&A_2:\quad
{\bm\alpha}''_1={\bm\alpha}_0,\quad
{\bm\alpha}''_2={\bm\alpha}_8.
\end{align}
The lattice originally spanned by ${\bm\alpha}_1,\cdots,{\bm\alpha}_8$
does not change even after we add ${\bm\alpha}_0$.
However, to remove ${\bm\alpha}_7$ and span it by
${\bm\alpha}_1,\cdots,{\bm\alpha}_6$ of $E_6$ and
${\bm\alpha}_0,{\bm\alpha}_8$ of $A_2$, it has to be
supplemented by
\begin{align}
{\bm\alpha}_7={\bm\omega}'_E+{\bm\omega}_A+\mbox{roots}.
\end{align}
Therefore, in the decomposition $E_8\to E_6\times A_2$, the root
lattice of $E_8$ is divided into the conjugacy classes
\begin{align}
\oplus_{k=0}^2k(1,1)=(0,0)\oplus(1,1)\oplus(2,2).
\label{E8E6A2}
\end{align}
Other decompositions of $E_8$ can be similarly found
as summarized in Table \ref{E8decomp}.

\begin{table}
\begin{center}
\begin{tabular}{c||c}
$A_1\times E_7$&$\oplus_{k=0}^1k(1,1)$\\\hline
$A_2\times E_6$&$\oplus_{k=0}^2k(1,1)$\\\hline
$A_3\times D_5$&$\oplus_{k=0}^3k(1,1)$\\\hline
$A_4\times A_4$&$\oplus_{k=0}^4k(1,2)$
\end{tabular}
\end{center}
\caption{Possible decompositions of the $E_8$ lattice.}
\label{E8decomp}
\end{table}

\begin{figure}[htb]
 \begin{center}
\scalebox{0.4}[0.4]{\includegraphics{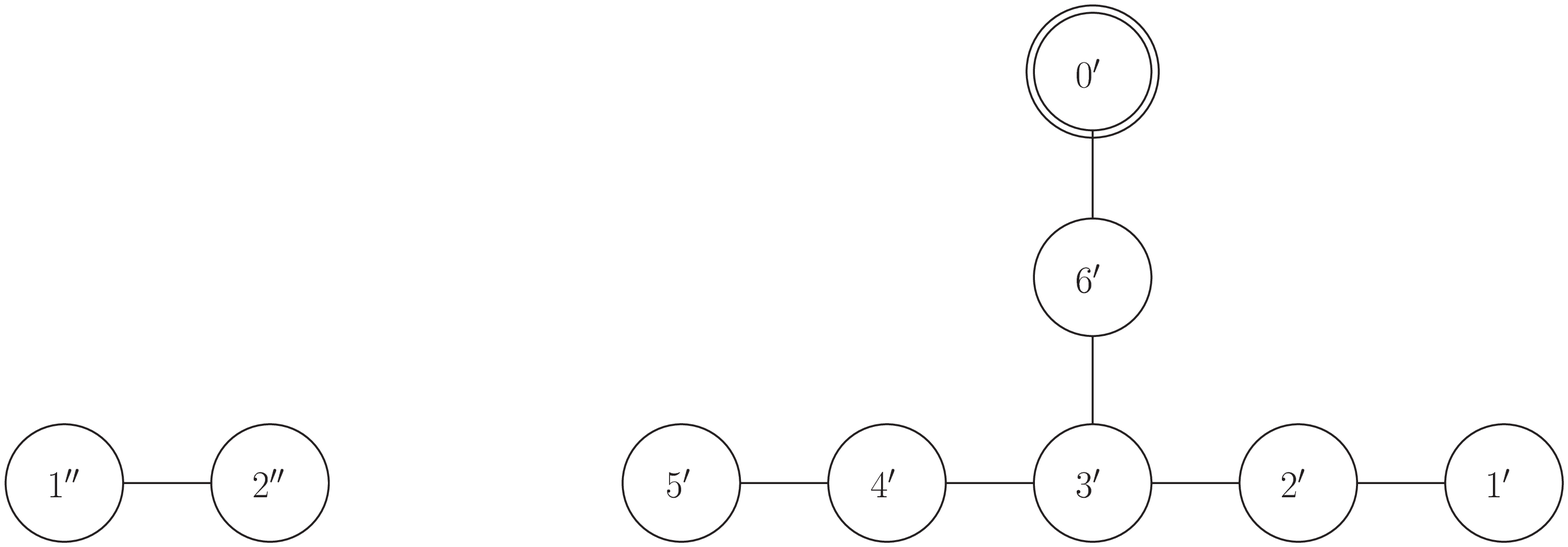}}
 \end{center}
 \caption{Dynkin diagrams of the $A_2$ group and $E_6$ group.}
 \label{fig:E6A2Dynkin}
 \end{figure}

\subsection{$E_6\to(A_2)^3$}
The next example of a decomposition is $E_6\to(A_2)^3$ 
as depicted in Figure \ref{fig:E6A2Dynkin}.
In this case we add the root ${\bm\alpha}'_0$ to obtain the
extended Dynkin diagram, where
\begin{align}
{\bm\alpha}'_0=-{\bm\alpha}'_1-2{\bm\alpha}'_2-3{\bm\alpha}'_3
-2{\bm\alpha}'_4-{\bm\alpha}'_5-2{\bm\alpha}'_6,
\end{align}
and remove the root ${\bm\alpha}'_3$.
Since the removed root ${\bm\alpha}'_3$ and the fundamental weight of
the $E_6$ algebra ${\bm\omega}'_E$ are decomposed in terms of the
$(A_2)^3$ lattice as
\begin{align}
{\bm\alpha}'_3
=1\cdot{\bm\omega}_A^{(1)}+1\cdot{\bm\omega}_A^{(2)}+1\cdot{\bm\omega}_A^{(3)}
+\mbox{roots},\quad
{\bm\omega}'_E
=0\cdot{\bm\omega}_A^{(1)}+1\cdot{\bm\omega}_A^{(2)}+2\cdot{\bm\omega}_A^{(3)}
+\mbox{roots},
\end{align}
we find that the various conjugacy classes are decomposed as
\begin{align}
0&\to 0\cdot(0,1,2)+\bigl[\oplus_{k=0}^2k(1,1,1)\bigr]
=(0,0,0)\oplus(1,1,1)\oplus(2,2,2),\nonumber\\
1&\to 1\cdot(0,1,2)+\bigl[\oplus_{k=0}^2k(1,1,1)\bigr]
=(0,1,2)\oplus(1,2,0)\oplus(2,0,1),\nonumber\\
2&\to 2\cdot(0,1,2)+\bigl[\oplus_{k=0}^2k(1,1,1)\bigr]
=(0,2,1)\oplus(2,1,0)\oplus(1,0,2).
\label{E6A2^3}
\end{align}

\subsection{$E_6\to D_4\times\widetilde A_2$}\label{lattice_decompD4A2}
Let us consider the decomposition
\begin{align}
E_6\to D_5\times U(1)\to D_4\times[U(1)]^2.
\end{align}
This time, by studying the decomposition carefully, we find that two pieces of
$U(1)$ actually take the form of an $A_2$ lattice, although the lattice spacing is
$\sqrt{2}$ times that of the original lattice (called $\widetilde A_2$ here),
and the decompositions of the various conjugacy classes are given by
\begin{align}
0&\to(0,\widetilde 0)
\oplus(v,\widetilde 0+\widetilde v)
\oplus(s,\widetilde 0+\widetilde s)
\oplus(c,\widetilde 0+\widetilde c),
\nonumber\\
1&\to(0,\widetilde 1)
\oplus(v,\widetilde 1+\widetilde v)
\oplus(s,\widetilde 1+\widetilde s)
\oplus(c,\widetilde 1+\widetilde c),
\nonumber\\
2&\to(0,\widetilde 2)
\oplus(v,\widetilde 2+\widetilde v)
\oplus(s,\widetilde 2+\widetilde s)
\oplus(c,\widetilde 2+\widetilde c).
\label{D4A2}
\end{align}
Here $v,s$ and $c$ of the $D_4$ lattice denote shifts by the fundamental
weights $\bm\omega^v$, $\bm\omega^s$ and $\bm\omega^c$, while
$\widetilde k$ in $\widetilde A_2$ means the shift by
$k\widetilde{\bm\omega}_1=k\sqrt{2}{\bm\omega}_1$ and
$\widetilde v, \widetilde s$ and $\widetilde c$ denote the shifts by the
vectors 
\begin{align}
\widetilde{\bm\omega}^v=\frac{\widetilde{\bm\alpha}_1}{2}
=\frac{\bm\alpha_1}{\sqrt{2}},\quad
\widetilde{\bm\omega}^s=\frac{\widetilde{\bm\alpha}_2}{2}
=\frac{\bm\alpha_2}{\sqrt{2}},\quad
\widetilde{\bm\omega}^c
=-\frac{\widetilde{\bm\alpha}_1+\widetilde{\bm\alpha}_2}{2}
=-\frac{\bm\alpha_1+\bm\alpha_2}{\sqrt{2}},
\end{align}
respectively.
A similar analysis shows that the $D_4$ lattice can be decomposed as
\begin{align}
D_4\to A_2\times\widetilde A_2.
\label{D4A2A2}
\end{align}

\end{document}